\documentclass[nofootinbib,aps,11pt]{revtex4-1}
\usepackage{graphicx}
\usepackage{subfigure} 
\usepackage{hyperref}
\usepackage{cancel}
\usepackage{amssymb}
\usepackage{textcomp}
\usepackage{amsmath}
\usepackage{bm}
\usepackage{times}
\usepackage{epsfig}
\usepackage{color}
\newcommand{\hc}{{\rm h.c.}}

\newcommand{\GeV}{{\rm GeV}}

\newcommand{\SM}{{\rm SM}}
\newcommand{\eq}{{\rm eq}}
\begin{document}
\title{ Shinning Light on Sterile Neutrino Portal Dark Matter from Cosmology and Collider }
\bigskip
\author{Ang Liu$^1$}
\author{Feng-Lan Shao$^1$}
\email{shaofl@mail.sdu.edu.cn}
\author{Zhi-Long Han$^2$}
\email{sps\_hanzl@ujn.edu.cn}
\author{Yi Jin$^2$}
\author{Honglei Li$^2$}
\affiliation{$^1$School of Physics and Physical Engineering, Qufu Normal University, Qufu, Shandong 273165, China\\
$^2$School of Physics and Technology, University of Jinan, Jinan, Shandong 250022, China}
\date{\today}
\begin{abstract}
 Provided the dark sector consisted of a dark scalar $\phi$ and a dark fermion $\chi$ under an exact $Z_2$ symmetry, the sterile neutrino $N$ can act as the messenger between the dark sector and standard model via the Yukawa coupling $\lambda_{ds} \bar{\chi}\phi N$. In this paper, we focus on the specific scenario $m_N>m_\phi+m_\chi$ with $\chi$ being a FIMP dark matter. The decay width of dark scalar $\phi$ is doubly suppressed by the smallness of Yukawa coupling $\lambda_{ds}$ and mixing angle $\theta$. The delayed decay $\phi\to\chi\nu$ will have a great impact on cosmological observables such as the Big Bang Nucleosynthesis, the Cosmic Microwave Background anisotropy power spectra, the effective number of relativistic neutrino species $N_{\rm eff}$ and the energetic neutrino flux. Meanwhile, the sterile neutrino can generate displaced vertex signature at colliders when $m_N<m_W$. The dark scalar $\phi$ will also induce measurable Higgs invisible decay for relatively large quartic coupling. A comprehensive analysis of constraints from cosmology and collider is performed in this paper. We find that almost the whole parameter space with $m_N<m_W$ is under the reach of future experiments.
 
\end{abstract}
\maketitle

\section{Introduction}
The origin of tiny neutrino mass and the existence of particle dark matter (DM) are the two concrete pieces of evidence for new physics beyond the standard model. Appealing pathways that connected these two parts together have been extensively studied in 
Refs.~\cite{Krauss:2002px,Asaka:2005an,Ma:2006km,Aoki:2008av,Cai:2017jrq}. The most economical way to explain the tiny neutrino mass is by introducing sterile neutrino $N$~\cite {Minkowski:1977sc,Mohapatra:1979ia}. Although a quite high scale $N$ ($m_N\gtrsim10^9$ GeV) is required by type-I seesaw and leptogenesis \cite{Fukugita:1986hr},  light sterile neutrino in the range of eV to TeV scale is also well studied \cite{Dasgupta:2021ies,Abdullahi:2022jlv}.

A light sterile neutrino could be directly produced at colliders ~\cite{Gorbunov:2007ak,Atre:2009rg,Deppisch:2015qwa}, which will lead to the distinct lepton number violation signature \cite{Cai:2017mow}. Meanwhile, if sterile neutrino $N$ is lighter than $W$ boson, the decay width of sterile neutrino $\Gamma_N$ is suppressed by the three-body phase space. Then $N$ becomes long-lived and leads to the displaced vertex signature \cite{Helo:2013esa,Alimena:2019zri}. This  signature is very promising to probe the mixing angle $\theta$ between the light and sterile neutrinos at present and future colliders \cite{Abdullahi:2022jlv}. Thus we focus on light sterile neutrino above the GeV scale in this paper.

Recently, the sterile neutrino $N$ as a portal to the dark sector via the Yukawa coupling $\lambda_{ds} \bar{\chi}\phi N$ is receiving increasing interest \cite{Escudero:2016tzx,Escudero:2016ksa,Coito:2022kif,Coy:2022xfj,Li:2022xjx}. In this paper, we consider the fermion singlet $\chi$ as dark matter. For a sizable coupling $\lambda_{ds}$, the secluded channel $\bar{\chi}\chi\to NN$ is important to obtain correct relic density, which is also observable at indirect detection experiments \cite{Campos:2017odj,Batell:2017rol,Folgado:2018qlv}. The DM-nucleon scattering cross section is one-loop suppressed, thus is easy to escape the tight direct detection limits. On the other hand, correct relic density can also be obtained for tiny via the freeze-in mechanism \cite{Bandyopadhyay:2020qpn,Cheng:2020gut,Falkowski:2017uya,Liu:2020mxj,Chang:2021ose}. Although the DM $\chi$ is hard to detect at canonical direct and indirect detection experiments, the delayed decay of dark scalar $\phi$ into light neutrinos will affect the Big Bang Nucleosynthesis (BBN) predictions, the Cosmic Microwave Background (CMB) anisotropy power spectra, the effective number of relativistic neutrino species $N_{\rm eff}$ and the energetic neutrino spectra observed today \cite{Adams:1998nr,Chen:2003gz,Boyarsky:2021yoh,Liu:2022rst}.

Provided the mass spectrum $m_N>m_\phi+m_\chi$, then the FIMP DM $\chi$ is produced from the out-of-equilibrium decay  $N\to \chi \phi$ \cite{Barman:2022scg}. Due to mixing between the light and sterile neutrinos, the dark scalar $\phi$ further decays via $\phi\to \chi\nu $. The decay width of dark scalar $\Gamma_\phi\simeq \lambda_{ds}^2 \theta^2 m_\phi/8\pi$ is heavily suppressed by the smallness of $\lambda_{ds}$ and $\theta$, therefore $\phi$ is a long-lived particle. For instance, with $\lambda_{ds}\sim10^{-11}$, $\theta\sim10^{-7}$, and $m_\phi\sim10^3$~GeV, we have the lifetime $\tau_\phi=1/\Gamma_\phi\sim10^{10}$ s. Such a long lifetime conflicts with constraints from CMB and BBN \cite{Hambye:2021moy}. One possible way to avoid the cosmological constraints is making $\phi$ a decaying DM via even smaller $\lambda_{ds}$ or $\theta$ \cite{Coy:2021sse}. Another pathway is assuming $m_\phi$ lighter than 200 GeV, so that the fraction of electromagnetic energy injected into the plasma is tiny.

In this paper, we perform a comprehensive analysis of freeze-in sterile neutrino portal dark matter under constraints from cosmology and colliders. We focus on the specific scenario $m_N>m_\phi+m_\chi$, which leads to $N\to \chi \phi$ followed by the delayed decay $\phi\to \chi\nu$. Since the Yukawa coupling $\lambda_{ds}$ is determined by relic density, the mixing angle $\theta$ plays an important role in the detection of this model. For a relatively larger mixing angle $\theta$, displaced vertex signature from sterile neutrino $N$ decay is observable at colliders. On the other hand, a relatively smaller mixing angle $\theta$ leads to the dark scalar $\phi$ long-lived, which has a great impact on cosmological observable. Therefore, complementary constraints are expected from cosmology and colliders.

The structure of this paper is organized as follows. In Sec. \ref{SEC:TM}, we briefly introduce the sterile neutrino portal DM model. The calculation of  DM relic density  is discussed in Sec. \ref{SEC:RD}. Cosmological constraints from CMB, BBN, $N_\text{eff}$ and neutrino flux on the long-lived dark scalar $\phi$ are considered in Sec. \ref{SEC:CN}.  Then we discuss the collider signatures as displaced vertex and Higgs invisible decay in Sec. \ref{SEC:CSN}. A scanning of the parameter space under combined constraints from cosmology and collider is performed in Sec.~\ref{SEC:CB}. Finally, we summarize our results in Sec. \ref{SEC:CL}

\section{The Model}\label{SEC:TM}

We consider a simple extension of the SM with sterile neutrino $N$ and a dark sector. Besides providing masses for SM neutrinos through the type-I seesaw mechanism, the sterile neutrino $N$ also interacts with particles in the dark sector to ensure the production of DM. One  scalar singlet $\phi$ and one fermion singlet $\chi$ are presumed in the dark sector, which is charged under the $Z_2$ symmetry. In this paper, we assume $m_\chi<m_\phi$, so that the DM candidate is the fermion singlet $\chi$.

The corresponding scalar potential with unbroken $Z_2$ symmetry could be denoted as
\begin{eqnarray} \label{sp}
	V & = & -\mu_{\Phi}^2 \Phi^\dag \Phi
	+\frac{\mu_\phi^{2}}{2} \phi^2+\frac{\lambda_1}{2} (\Phi^\dag \Phi)^2+\frac{\lambda_2}{4} \phi^4+\lambda_3 \phi^2(\Phi^\dag\Phi),
\end{eqnarray}
where $\Phi$ is the standard Higgs doublet. After the spontaneous symmetry breaking, we have one physical Higgs boson $h$ and one dark scalar $\phi$ with mass $m_\phi^2=\mu^2_\phi+\lambda_3 v^2$. Due to the introduction of new particles,
the new  Yukawa interaction and mass terms can be written as 
\begin{equation}\label{yuk}
	-\mathcal{L}_Y\supset y\overline{L} \widetilde{\Phi} N  +\lambda_{ds} \bar{\chi}\phi N  +\frac{1}{2}\overline{N ^c}m_{N } N  + m_\chi \bar{\chi} \chi+ \hc,
\end{equation}
where $\widetilde{\Phi}=i\sigma_2 \Phi^*$. In this paper, we assume $\lambda_{ds}\ll1$ to make $\chi$ a FIMP dark matter. The resulting light neutrino mass is
\begin{equation}
	m_\nu=-\frac{v^2}{2} y\, m_N^{-1} y^T.
\end{equation}
In principle, the mixing angle between the light and sterile neutrino can be quantified as
\begin{equation}\label{Eqn:SS}
	\theta=\frac{ yv}{\sqrt{2}m_N}\sim\sqrt{m_\nu/m_N}.
\end{equation}
Typically, the mixing angle $\theta\sim10^{-6}$ is predicted by the seesaw relation with $m_\nu\sim0.1$ eV and $m_N\sim100$~GeV. Such a small mixing angle is beyond the reach of future colliders\cite{Abdullahi:2022jlv}. In this paper, we take $\theta$ as a free parameter. Large mixing angle is possible in low scale models, such as inverse seesaw \cite{Mohapatra:1986bd,Mohapatra:1986aw} and linear seesaw \cite{Wyler:1982dd,Akhmedov:1995ip,Akhmedov:1995vm}.

\section{Relic Density} \label{SEC:RD} 
In this paper, we focus on the specific scenario $m_N>m_\phi+m_\chi$.
As the FIMP DM candidate, the production of fermion $\chi$ is through the direct decay $N\to \chi \phi$ followed by the delayed decay $\phi\to \chi \nu$. For the dark scalar $\phi$, additional contributions should be considered due to the Higgs portal interaction $\lambda_3 \phi^2 (\Phi^\dag \Phi)$. For large enough $\lambda_{3}$, the abundance of dark scalar $Y_\phi$ is dominantly determined by the annihilation process $\phi\phi\to \SM$ via the freeze-out mechanism. On the other hand, a tiny $\lambda_{3}$ will lead to an additional contribution of $Y_\phi$ by the process $\SM \to \phi\phi$ via the freeze-in mechanism. Both the WIMP and FIMP scalar scenarios will be considered in this paper.

\begin{figure}
	\begin{center}
		\includegraphics[width=0.45\linewidth]{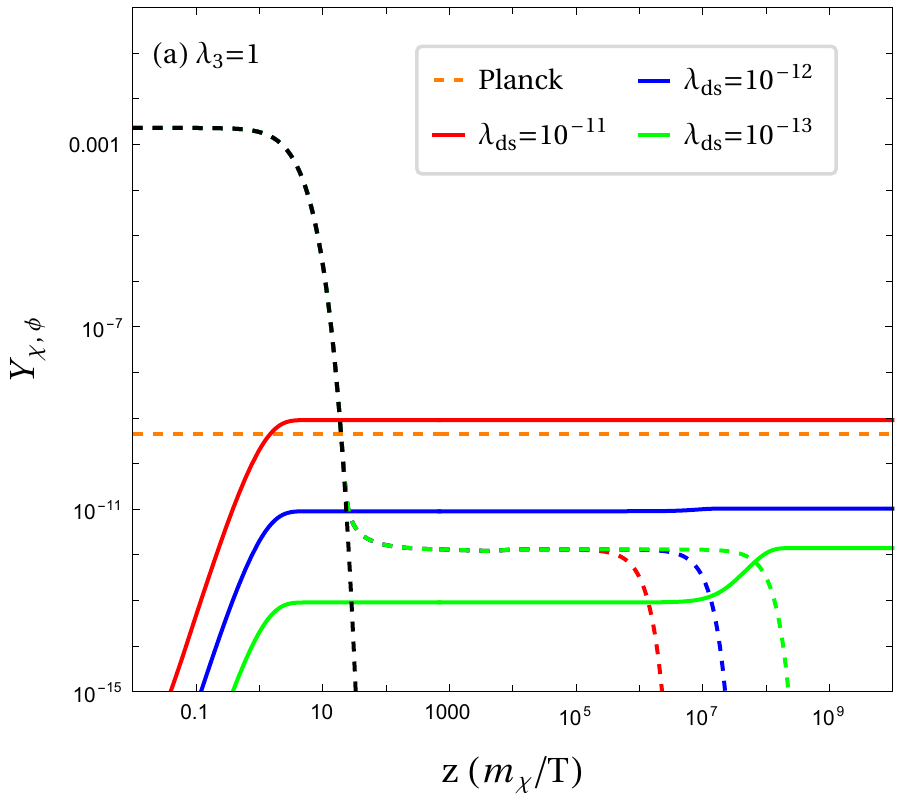}
		\includegraphics[width=0.45\linewidth]{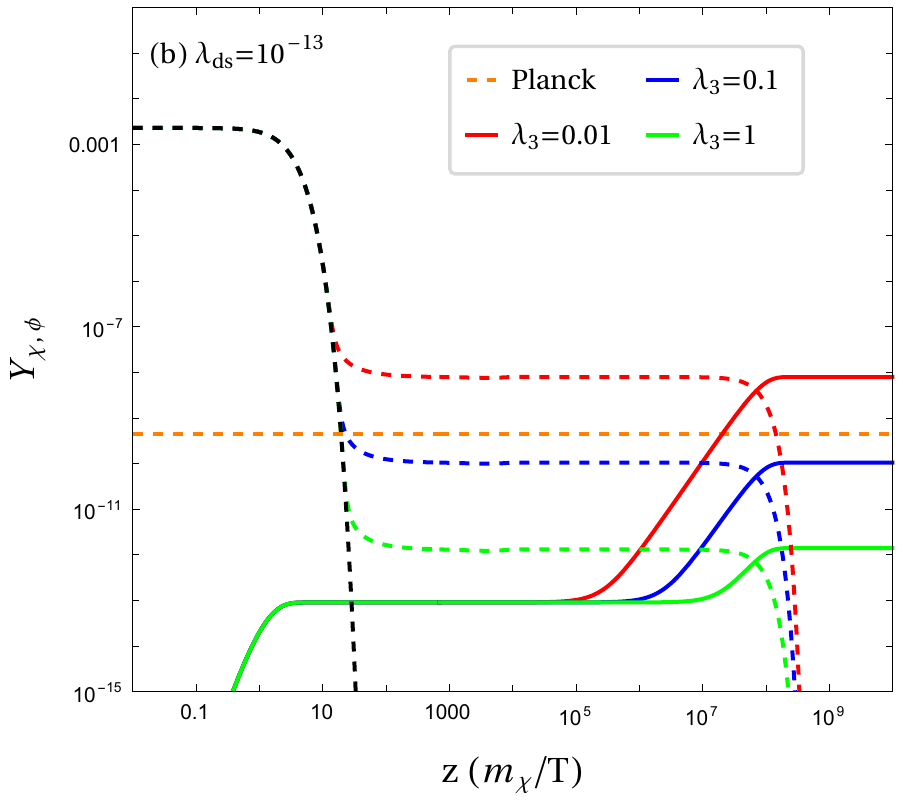}
		\includegraphics[width=0.45\linewidth]{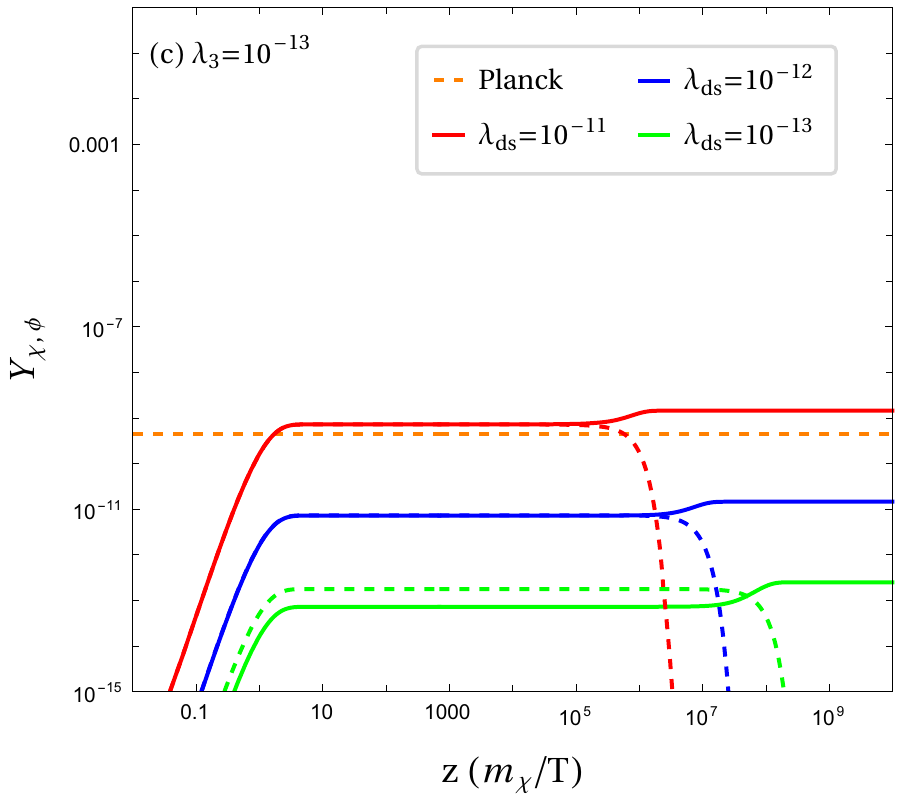}
		\includegraphics[width=0.45\linewidth]{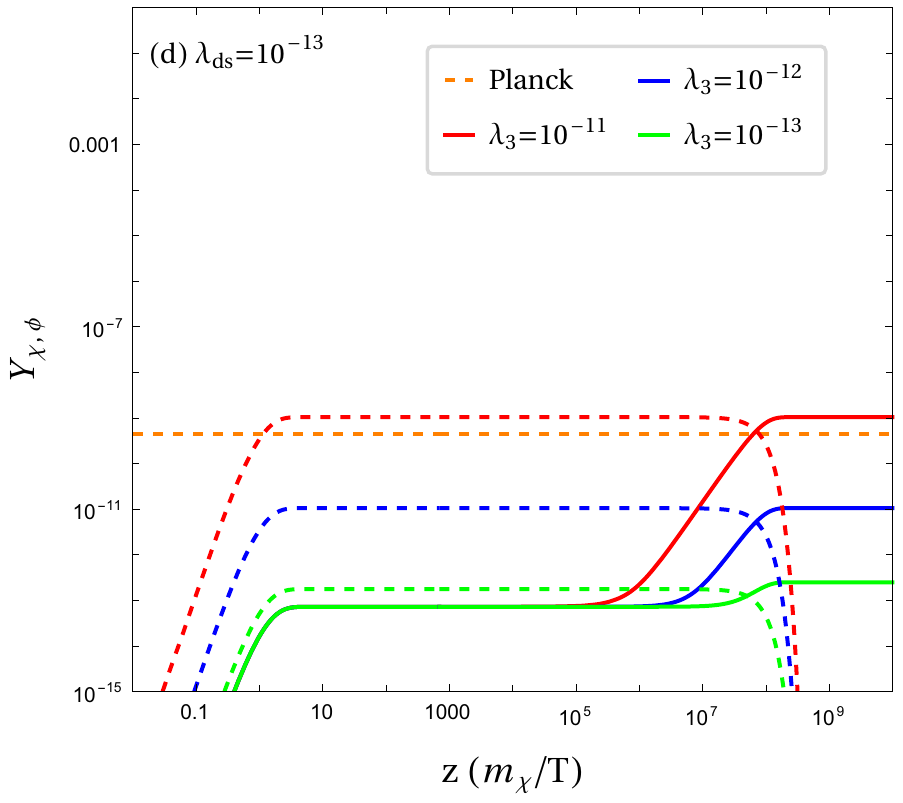}
	\end{center}
	\caption{ The evolution of dark sector abundances. The red, blue and green solid lines represent the evolution of DM $\chi$, and the corresponding dotted lines express the evolution of  dark scalar $\phi$. Subfigures (a) and (b) are WIMP scalar scenario with $\lambda_{3}\sim\mathcal{O}(1)$. Subfigure (c) and (d) are FIMP scalar scenario with $\lambda_{3}\ll\mathcal{O}(1)$. The orange horizontal lines  are the Planck observed relic density $\Omega_\text{DM}h^2=0.12$ \cite{Planck:2018vyg} for 
		$m_\text{DM}=1$ GeV. The black dashed lines describe the evolution of $Y_\phi^\text{eq}$.}
	\label{FIG:DM}
\end{figure}

The Boltzmann equations describing the evolution of dark sector abundances are:
\begin{eqnarray}\label{f-y}
	\frac{dY_\phi}{dz} &= &
	k^{\star}z\tilde{\Gamma}_{N\to\phi\chi}\left(Y_{N}^\eq-\frac{Y_{N}^\eq}{Y_{\phi}^\eq Y_{\chi}^\eq}Y_\phi Y_\chi\right)-k^{\star}z \tilde{\Gamma}_{\phi\to\chi \nu}\left(Y_{\phi}-\frac{Y_{\phi}^\eq}{Y_{\chi}^\eq}Y_\chi\right)
	\nonumber \\
	&+&\frac{k}{z^2} \left \langle \sigma v \right \rangle_{\SM\to\phi\phi}\left((Y_{\SM}^\eq)^2-\left(\frac{Y_{\SM}^\eq}{Y_{\phi}^\eq}\right)^2 Y_{\phi}^{2}\right) ,\\
	\frac{dY_\chi}{dz} &= & k^{\star}z\tilde{\Gamma}_{N\to\phi\chi}\left(Y_{N}^\eq-\frac{Y_{N}^\eq}{Y_{\phi}^\eq Y_{\chi}^\eq}Y_\phi Y_\chi\right)+k^{\star}z \tilde{\Gamma}_{\phi\to\chi \nu}\left(Y_{\phi}-\frac{Y_{\phi}^\eq}{Y_{\chi}^\eq}Y_\chi\right),	
\end{eqnarray}
where we use the definition $z\equiv m_\chi/T$, and $T$ is the temperature. The parameters $k$ and $k^{\star}$ are denoted as $k=\sqrt{\pi g_\star/45}m_\chi M_{Pl}$ and $k^{*}=\sqrt{45/4\pi^{3}g_{\star}}M_{Pl}/m_\chi^2$ respectively, where $g_{\star}$ is the effective number of degrees of freedom of the relativistic species and $M_{Pl}=1.2 \times 10^{19}$ GeV is the Planck mass. In the above Boltzmann equations, contributions from scattering processes as $\phi\phi\to \chi\chi$, $NN\to \chi\chi$, $NN\to\phi\phi$, $h\nu\to \phi\chi$ and $hN\to \phi\chi$ are not considered, since the corresponding cross sections are suppressed \cite{Bandyopadhyay:2020qpn,Cheng:2020gut}. We use micrOMEGAs \cite{Belanger:2013oya} to calculate the thermal average cross sections $\langle \sigma v \rangle$. The thermal decay width $\tilde{\Gamma}_i$ is defined as $\Gamma_i \mathcal{K}_1/\mathcal{K}_2$ with $\mathcal{K}_{1,2}$ being the first and second modified Bessel Function of the second kind. 
Corresponding decay widths are given by 
\begin{eqnarray}
	\Gamma_{{N\rightarrow}{\phi\chi}}&=&\frac {\lambda_{ds}^{2}}{16\pi}\frac{(m_N+m_\chi)^2-m_\phi^2}{m_N^3}{\lambda^{1/2}(m_N^2,m_\phi^2,m_\chi^2)}, \\
	\Gamma_{\phi\to \chi \nu} &=& \frac {\lambda_{ds}^{2}\theta^2\, m_\phi}{8\pi}{\left(\frac{m_\phi^2-m_\chi^2}{m_\phi^2}\right)^2}.
\end{eqnarray}
The kinematic function $\lambda(a,b,c)$ is
\begin{equation} 
	\lambda(a,b,c)=a^2+b^2+c^2-2ab-2ac-2bc.
\end{equation}

The evolution of dark sector abundances for various benchmark scenarios is shown in Fig.~\ref{FIG:DM}.  We fix $m_\chi=1$ GeV, $m_\phi=10$  GeV and $m_N=50$ GeV for illustration. In panel (a) of Fig.~\ref{FIG:DM}, we choose $\lambda_{3}=1$ while varying $\lambda_{ds}$. Such large $\lambda_{3}$ leads to $Y_\phi\sim10^{-12}$, which is far below the Planck required value $Y_\chi\simeq4.4\times10^{-10}$. In this case, we need DM $\chi$ generating from direct decay $N\to \phi \chi$ to be the dominant channel with $\lambda_{ds}\sim10^{-11}$. It is clear that varying $\lambda_{ds}$ also change the lifetime of $\phi$. For the WIMP scalar scenario, the dark scalar $\phi$ is produced via freeze-out. Therefore, the smaller the Higgs portal coupling $\lambda_{3}$ is, the larger the scalar abundance $Y_\phi$ will be. Fixing $\lambda_{ds}=10^{-13}$, correct relic density is also possible from delayed decay $\phi\to \nu \chi$ with $\lambda_{3}\sim 0.1$ as shown in panel (b) of Fig.~\ref{FIG:DM}. 

For the FIMP scalar scenario, the direct decay $N\to \phi \chi$ will result in equal abundances $Y_\phi=Y_\chi$ when $\lambda_{ds}\gg \lambda_{3}$ at the very beginning. Changing $\lambda_{ds}$ not only affects the relic density, but also the lifetime of $\phi$. According to panel (c) of Fig.~\ref{FIG:DM}, the correct relic density is obtained with $\lambda_{ds}\sim10^{-11}$. In the opposite case with $\lambda_{3}\gg\lambda_{ds}$, the dark scalar is dominantly produced from Higgs portal interaction $h\to \phi\phi$. The delayed decay $\phi\to \nu\chi$ can generate observed relic density with $\lambda_{3}\sim10^{-11}$, which is depicted in panel (d) of Fig.~\ref{FIG:DM}. 

In summary, when the direct decay $N\to \phi \chi$ is the dominant channel, the coupling $\lambda_{ds}$ is fixed by the relic density. However, when the delayed decay $\phi\to \nu\chi$ is the dominant one, the relic density of DM $\chi$ is actually determined by the Higgs portal coupling $\lambda_3$. In this way, both the WIMP scalar with $\lambda_3\sim\mathcal{O}(10^{-1})$ and the FIMP scalar with $\lambda_3\sim\mathcal{O}(10^{-11})$ are viable. Notably, $Y_\phi$ before $\phi$ decay in the WIMP scalar scenario can be much smaller than $Y_\chi$ for large enough $\lambda_{3}$, while it is always larger than $Y_\chi$ in the FIMP scalar scenario due to additional contributions from ${\rm SM}\to\phi\phi$.

\section{Dark Scalar in Cosmology} \label{SEC:CN}

As mentioned above, the long-lived dark scalar $\phi$ will decay into DM $\chi$ and light neutrino $\nu$. The energetic neutrino and secondary particles from it lead to observable signature. In this section, we mainly study constraints from three aspects: cosmological probes from CMB and BBN, the effective number of relativistic neutrino species $N_{\rm eff}$ and the energetic neutrino flux observed today.

\subsection{Constraints from CMB and BBN}

\begin{figure}
	\begin{center}
		\includegraphics[width=0.45\linewidth]{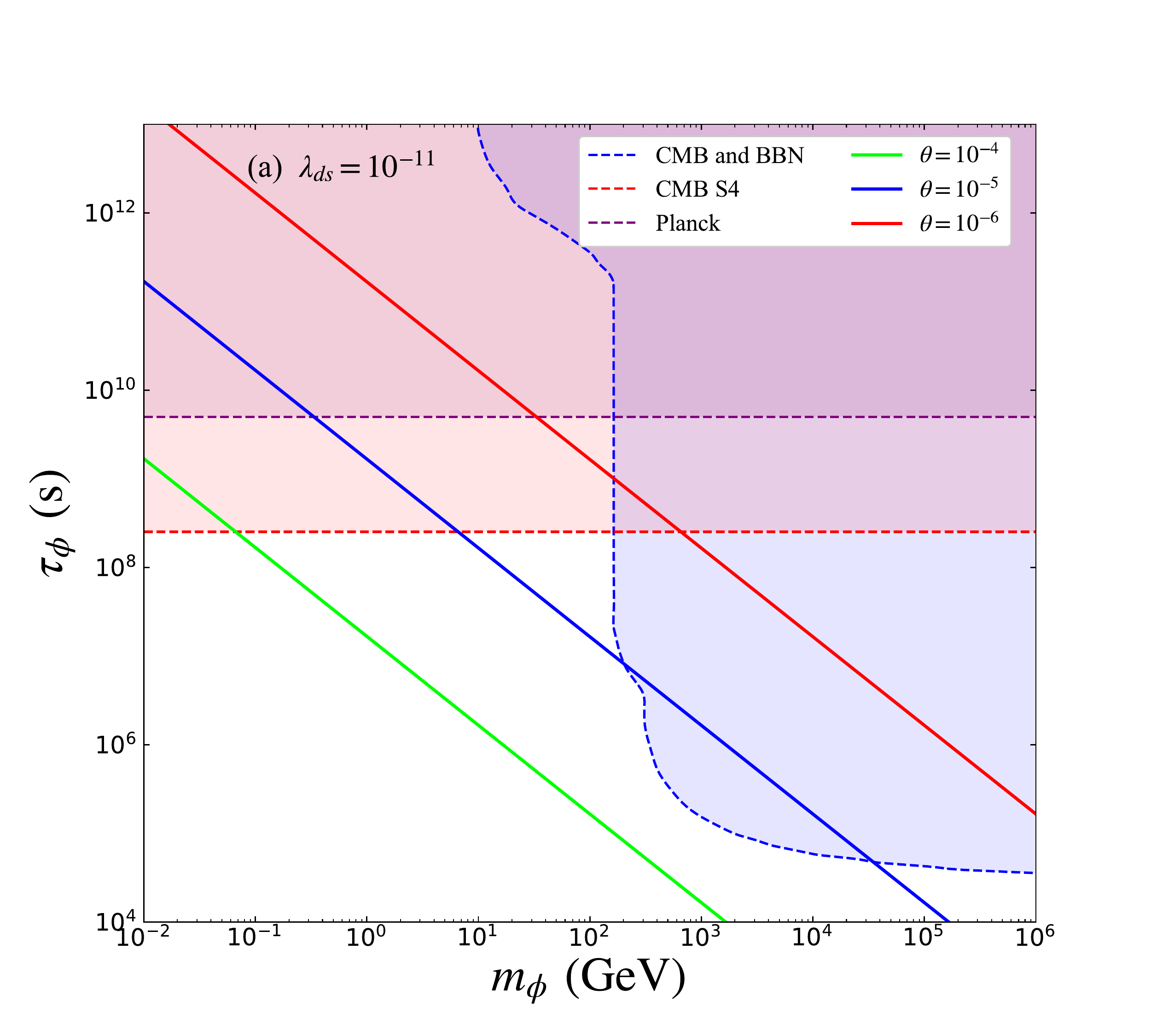}
		\includegraphics[width=0.45\linewidth]{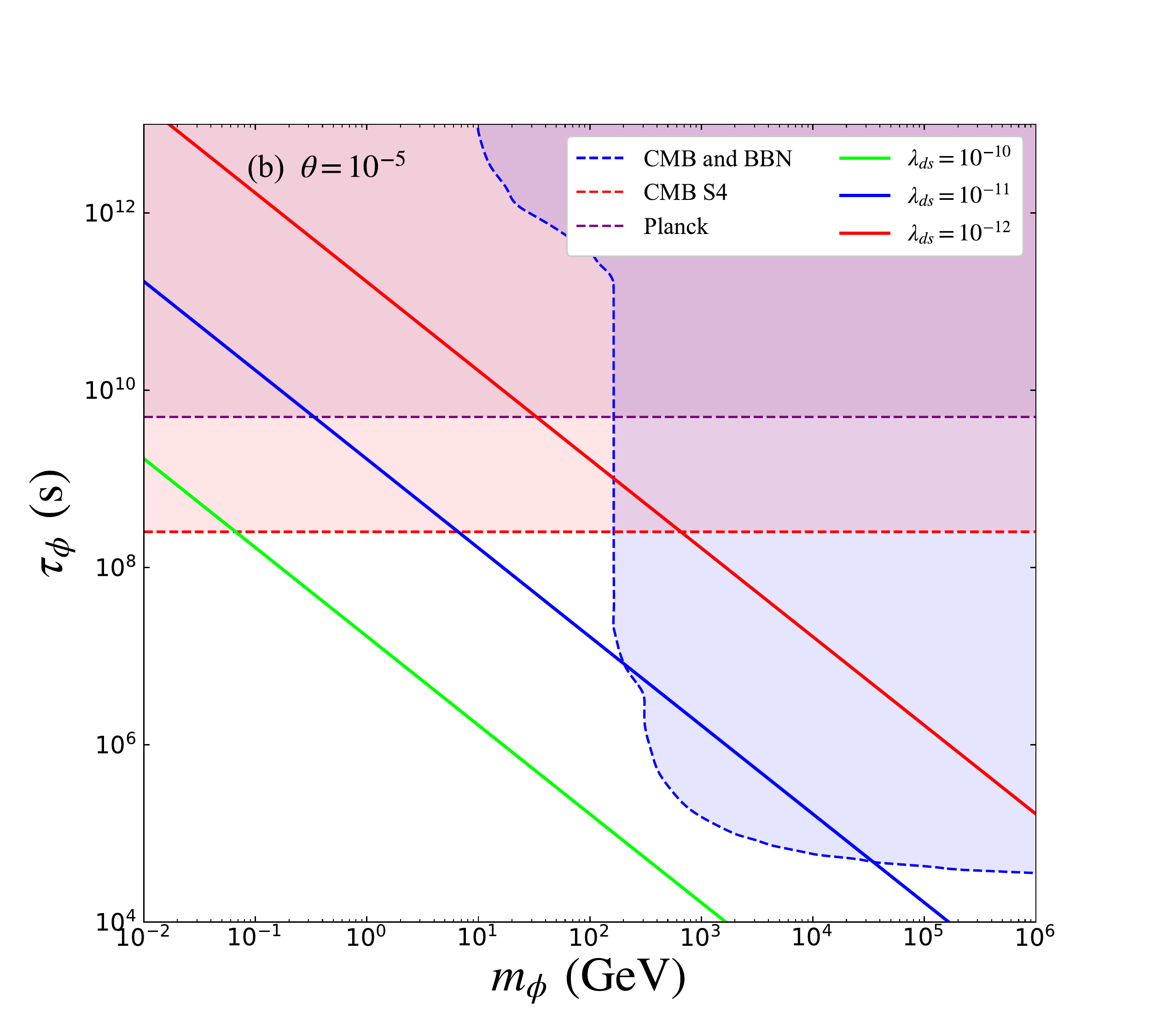}
	\end{center}
	\caption{The predicted lifetime of dark scalar $\tau_\phi$ for some benchmark scenarios. The blue region is excluded by CMB and BBN with $f_\phi=1$ ~\cite{Hambye:2021moy}. The purple region conflicts with the $N_{\rm eff}$ result by Planck, and the red region is the future reach of the CMB S4 experiment.}
	\label{FIG:CMB}
\end{figure}

For the cosmological constraints from CMB and BBN, we mainly consider the bounds discussed in Ref.~\cite{Hambye:2021moy}. The neutrinos produced by the delayed decay of dark scalar $\phi$ will emit secondary particles which occur in electromagnetic interactions. This electromagnetic material will affect electromagnetically cosmological probes, such as CMB anisotropies, CMB spectral distortions, and BBN photodisintegration. These effects are determined by the lifetime of dark scalar $\tau_\phi$ and the fractional abundance $f_\phi\equiv\Omega_{\phi}/\Omega_{\chi}$, where $\Omega_{\phi}$ is the abundance that  $\phi$ would have today if it was not decaying.

In Fig.~\ref{FIG:CMB}, we show the predicted lifetime $\tau_\phi$ as a function of $m_\phi$ with corresponding constraints from CMB, BBN and $N_{\rm eff}$. Detailed discussion on constraints from $N_{\rm eff}$ will be presented in the next subsection. In panel (a) of Fig.~\ref{FIG:CMB}, we have fixed $\lambda_{ds}=10^{-11}$, which corresponds to the scenario when DM is produced from direct decay $N\to \phi \chi$. For mixing angle $\theta=10^{-6}$, only the region around $m_\phi\sim100$ GeV is allowed by the current experiment. Increasing the value of $\theta$ will decrease the lifetime $\tau_\phi$, thus easier to satisfy experimental limits. For instance, the allowed region is extended to about $0.3\sim200$ GeV when $\theta=10^{-5}$, and the whole mass region above 0.01 GeV is allowed for $\theta=10^{-4}$ at present. In the future, the CMB S4 experiment can probe the region below 0.1 GeV for $\theta=10^{-4}$. In panel (b) of Fig.~\ref{FIG:CMB}, we have fixed $\theta=10^{-5}$. It is clear that varying $\lambda_{ds}$ leads to similar results as varying $\theta$. That is to say, the larger the coupling $\lambda_{ds}$ is, the smaller the lifetime $\tau_{\phi}$ is. On the other hand, we have to keep in mind that a too large coupling $\lambda_{ds}$ could lead to the production of DM over abundance. Notably, a much heavier dark scalar, e.g., $m_\phi\gtrsim10^5$ GeV with $\theta=10^{-5}$ and $\lambda_{ds}=10^{-11}$, is also possible to satisfy experimental limits. However, the requirement $m_N>m_\phi$ in this paper leads the sterile neutrinos far beyond collider search. So we do not consider this scenario.

\subsection{Constraints from $N_{\rm eff}$}

\begin{figure}
	\begin{center}
		\includegraphics[width=0.45\linewidth]{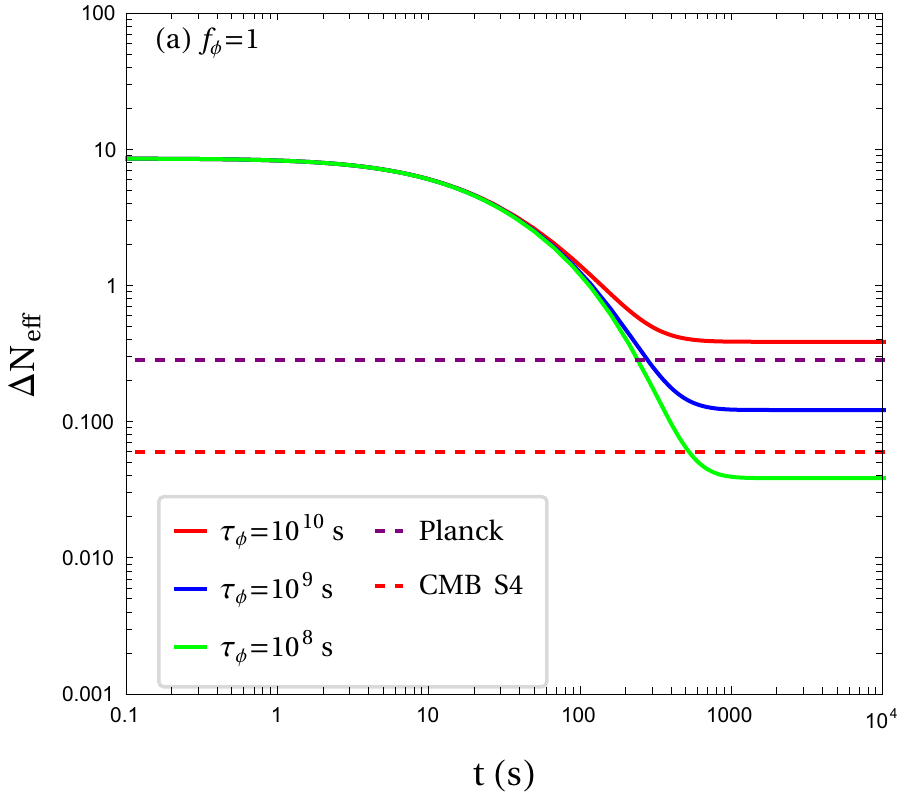}
		\includegraphics[width=0.45\linewidth]{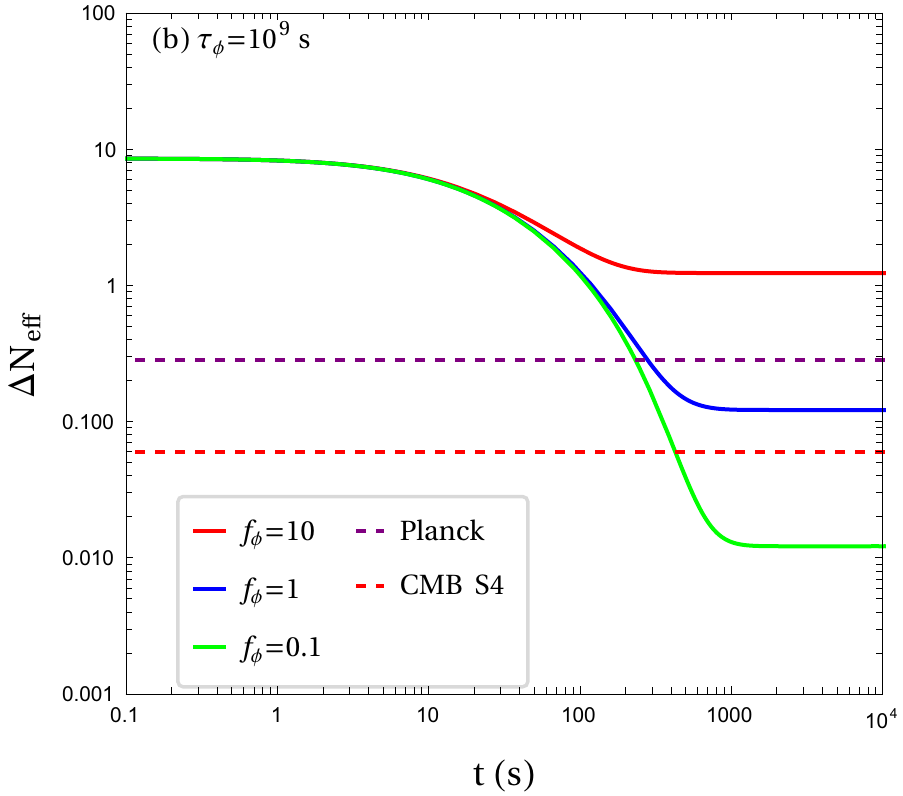}
	\end{center}
	\caption{The evolution of $\Delta N_{\rm eff}$ for some benchmark scenarios. The calculations are started at $T_\gamma=T_\nu=10$ MeV with the corresponding initial time  $t_0=\frac{1}{2H}|_{T=10 ~\rm MeV}$. The purple and red dashed lines represent the constraints of $\Delta N_{\rm eff}$ from current Planck~\cite{Planck:2018vyg} and future CMB S4~\cite{Abazajian:2019eic}, respectively. }
	\label{FIG:Neff}
\end{figure}

In our model, the delayed decay $\phi\to \chi \nu$ will increase
the effective number of relativistic neutrino species $N_{\rm eff}$.
The expression of $N_{\rm eff}$ can be written as:
\begin{equation}\label{Neff}
N_{\rm eff}=\frac{7}{8}\left(\frac{11}{4}\right)^{4/3}\left(\frac{\rho_\nu}{\rho_\gamma}\right)=3\left(\frac{11}{4}\right)^{4/3}\left(\frac{T_\nu}{T_\gamma}\right)^4,
\end{equation}
where $\rho_\nu$ and $\rho_\gamma$ represent the energy densities of active neutrinos and photons respectively, $T_\nu$ and $T_\gamma$ are their corresponding temperatures.
According to the evolution equations of $T_\nu$ and $T_\gamma$ in SM~\cite{EscuderoAbenza:2020cmq, Escudero:2018mvt}, the corresponding equations that conform to our model are modified as
 \begin{eqnarray}\label{Tnu}
 	\frac{dT_{\gamma}}{dt}&=&-\frac{4H\rho_{\gamma}+3H(\rho_e+p_e)+\frac{\delta\rho_{\nu_e}}{\delta t}+2\frac{\delta\rho_{\nu_\mu}}{\delta t}-\varepsilon\xi_{\rm EM}\frac{\rho_{\phi}}{\tau_{\phi}}}{\frac{\partial\rho_{\gamma}}{\partial T_{\gamma}}+\frac{\partial\rho_{e}}{\partial T_{\gamma}}},\\
 	\frac{dT_{\nu}}{dt}&=&-H T_{\nu}+\frac{\frac{\delta\rho_{\nu_e}}{\delta t}+2\frac{\delta\rho_{\nu_\mu}}{\delta t}+\varepsilon(1-\xi_{\rm EM})\frac{\rho_{\phi}}{\tau_{\phi}}}{3\frac{\partial\rho_{\nu}}{\partial T_{\nu}}}.
\end{eqnarray}
Here we assume that three flavor neutrinos have the same temperature. $\rho_{\gamma,e,\nu}$ are the energy densities of $\gamma$, $e$ and $\nu$. $\rho_{\phi}$ expresses the energy density of $\phi$ if it does not decay. $p_e$ is the pressure density of $e$. $\varepsilon=(m^2_{\phi}-m^2_{\chi})/2m^2_{\phi}$ denotes the fraction of the energy of $\phi$ that has been transferred to neutrinos~\cite{Blackadder:2014wpa}. $\xi_{\rm EM}$ represents the energy fraction that the neutrinos inject into electromagnetic plasma. Provided $m_\phi\lesssim100$~GeV, vanishing $\xi_{\rm EM}$ is assumed in our calculation \cite{Hambye:2021moy}. The last neutrino-electron energy density transfer rate $\frac{\delta\rho_{\nu}}{\delta t}$ can be obtained in Refs.~\cite{Escudero:2018mvt, EscuderoAbenza:2020cmq}.

The results of $\Delta N_{\rm eff}=N_{\rm eff}-N^{\rm SM}_{\rm eff}$ as a function of cosmic time $t$ are shown in Fig.~\ref{FIG:Neff}, where $N^{\rm SM}_{\rm eff}=3.045$ is considered~\cite{Mangano:2005cc, Grohs:2015tfy, deSalas:2016ztq}. 
Here we fix $m_{\chi}=10$ GeV and $m_{\phi}=50$ GeV for illustration.  For $f_\phi=1$, a lifetime $\tau_{\phi}=10^{10}$ s has been excluded by Planck measurement. Meanwhile, a lifetime $\tau_{\phi}=10^{9}$ s with $f_\phi=1$ leads to $\Delta N_{\rm eff}=0.12$, which is within the reach of future CMB S4 experiment. And a lifetime down to about $10^8$ s will beyond the scope of future limit. If fixing $\tau_{\phi}=10^9$ s, it is clear that $f_\phi=10$ predicts a too large value of $\Delta N_{\rm eff}$. To avoid future CMB S4 limit, we need $f_\phi\sim\mathcal{O}(0.1)$ with $\tau_{\phi}=10^9$ s.
Approximately, constraints from $N_{\rm eff}$ are proportional to the product of $\tau_{\phi}$ and $f_{\phi}$.  Currently, the Planck result requires $f_{\phi}^2\tau_{\phi}\lesssim5\times10^9$ s, and the future CMB S4 could push to $f_{\phi}^2\tau_{\phi}\lesssim2.5\times10^8$ s. The corresponding exclusion limits are also shown in Fig.~\ref{FIG:CSM}.

\subsection{Constraints from Neutrino Flux}
The energetic neutrinos induced by the delayed decay of $\phi$ might be probed by neutrino experiments. The neutrino flux at present is calculated as~\cite{Bandyopadhyay:2020qpn},

\begin{equation}
	\Phi_{\rm cos}\equiv E_\nu^2 \frac{d\varphi}{dE_\nu}=E_\nu\left(\frac{n_{\phi}}{\tau_{\phi}}\right)\left(\frac{e^{-t(x)/\tau_{\phi}}}{H(x)}\right)\theta^{'}(x),
\end{equation}
where $E_\nu$ represents the observed neutrino energy, $d\varphi/dE_\nu$ is the predicted neutrino flux, $n_{\phi}$ is the number density of $\phi$ if it is stable, $\theta^{'}(x)$ is the Heaviside theta function. The cosmic time $t(x)$ at red-shift $1+x$ and the Hubble parameter $H(x)$ in the standard cosmology are given by
\begin{eqnarray}\label{nf2}
	t(x)&\approx& \frac{4}{3H_0}\left(\frac{\Omega_{\rm r}^{3/2}}{\Omega_{\rm m}^{2}}\right)\left(1-\left(1-\frac{\Omega_{\rm m}}{2(1+x)\Omega_{\rm r}}\right)\sqrt{1+\frac{\Omega_{\rm m}}{(1+x)\Omega_{\rm r}}}\right),  \\
	H(x)&=&H_0\sqrt{\Omega_\Lambda+(1+x)^3\Omega_{\rm m}+(1+x)^4\Omega_{\rm r}},
\end{eqnarray}
where $x=E_0/E_\nu-1$ with initial energy $E_0=(m_{\phi}^2-m_{\chi}^2)/2m_{\phi}$, the Hubble constant $H_0=100h~\rm{km/s/Mpc}$ with $h=0.6727$ \cite{Planck:2018vyg}. The dark energy, matter and radiation (CMB photons and neutrinos) fractions are $\Omega_\Lambda=0.6846, \Omega_{\rm m}=0.315$ and $\Omega_{\rm r}=9.265\times10^{-5}$. 

\begin{figure}
	\begin{center}
		\includegraphics[width=0.45\linewidth]{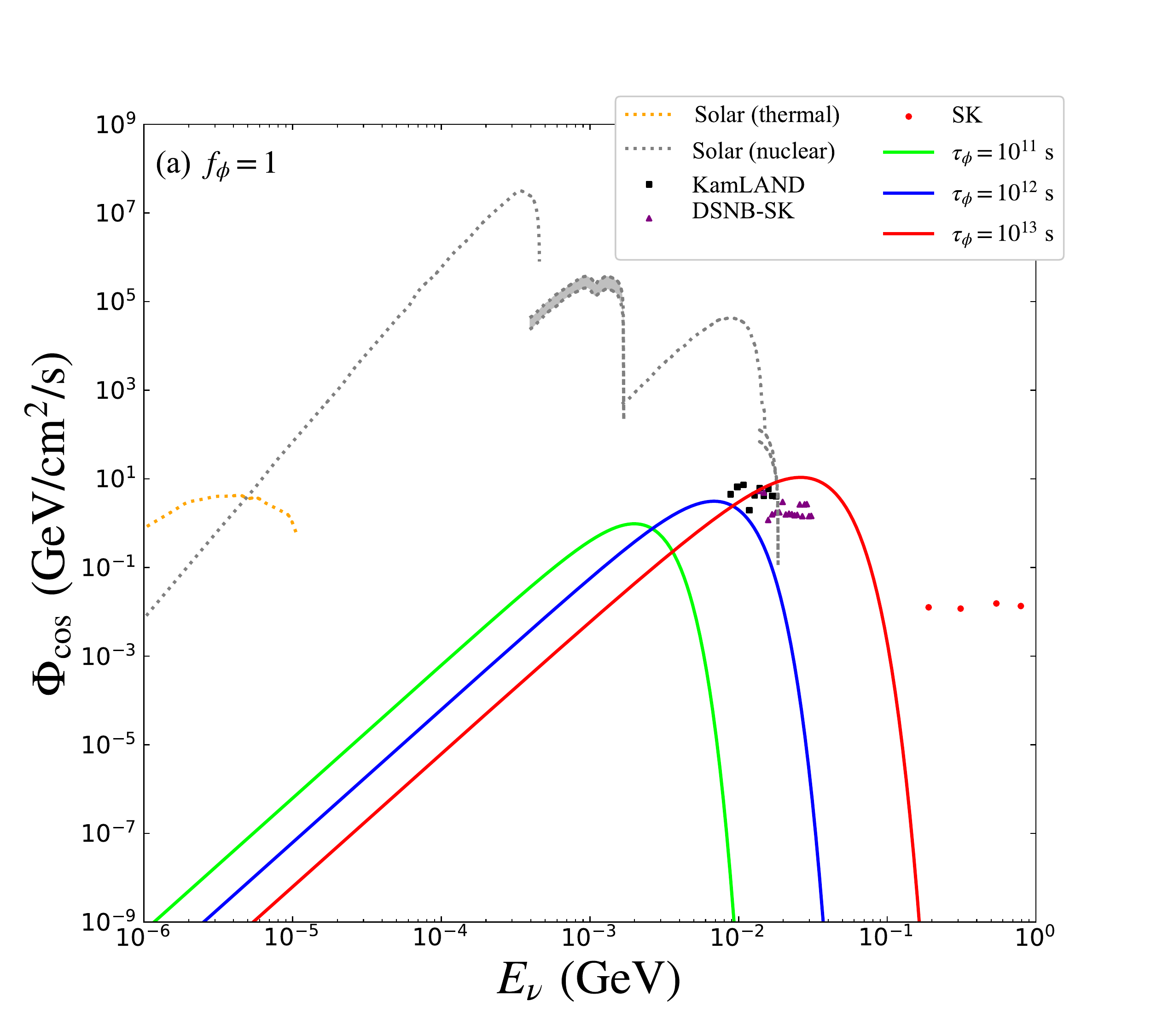}
		\includegraphics[width=0.45\linewidth]{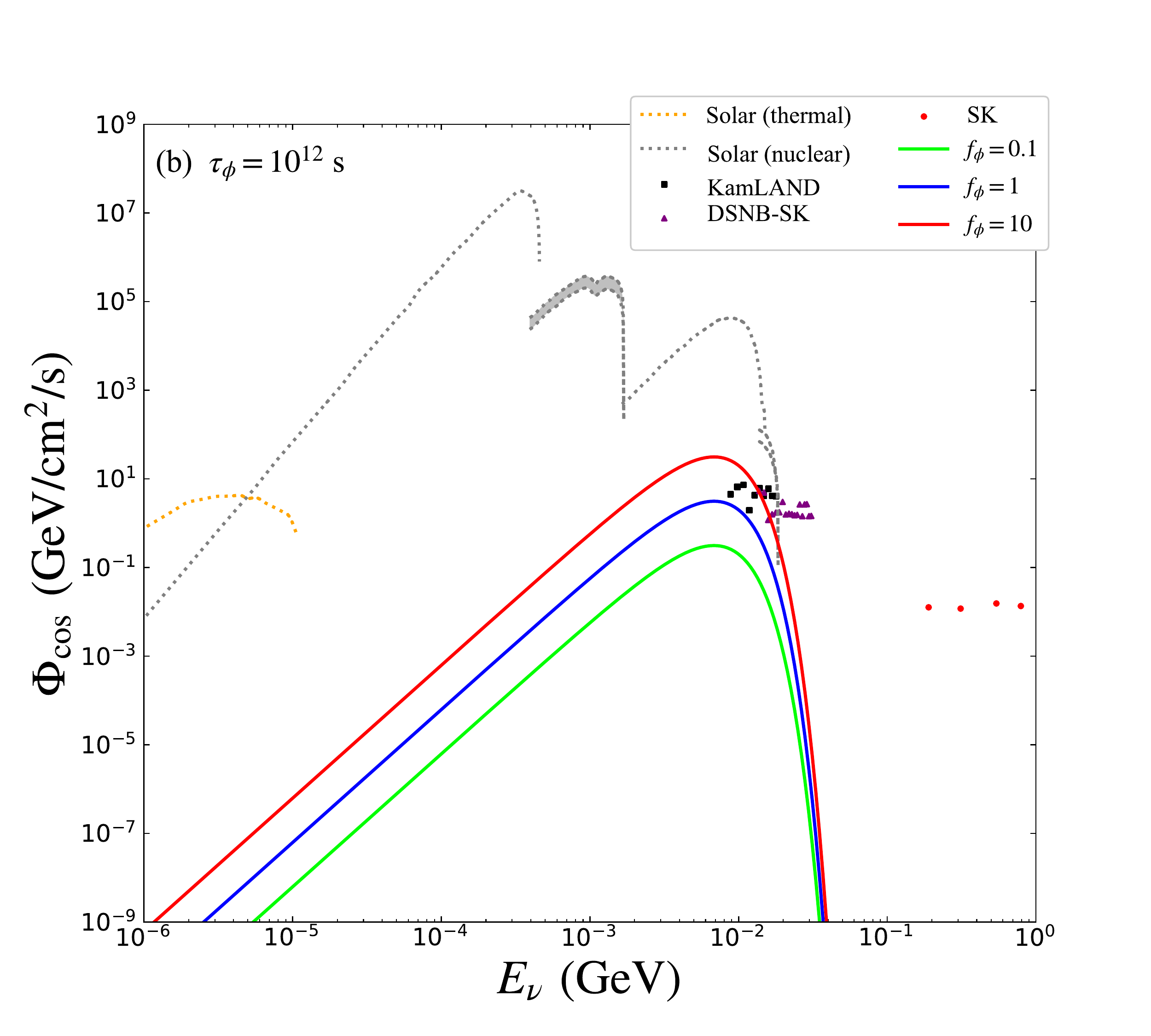}
	\end{center}
	\caption{The predicted neutrino fluxes at present for various scenarios.  The orange and gray dotted lines are the thermal and nuclear solar neutrino flux \cite{Vitagliano:2019yzm}. The black squares and purple triangles represent the diffuse supernova neutrino background (DSNB) flux measured at the KamLAND \cite{KamLAND:2011bnd} and SK \cite{Super-Kamiokande:2013ufi}, respectively. The red points are the atmospheric neutrino data from SK \cite{Super-Kamiokande:2015qek}.}
	\label{FIG:vflux}
\end{figure}
The neutrino fluxes for different parameters are shown in Fig.~\ref{FIG:vflux}. We fix $m_{\chi}=10$ GeV, $m_{\phi}=50$ GeV in the calculation. For fixed $f_\phi$, the observed energy $E_\nu$ and the maximum neutrino flux increase with the growth of $\tau_{\phi}$. It is clear that $\tau_{\phi}\gtrsim10^{13}$ s  with $f_{\phi}=1$ is excluded by KamLAND and SK data. Meanwhile, a larger $f_\phi$ will also increase the neutrino flux for fixed $\tau_\phi$.  The predicted neutrino energy is usually less than 100 MeV with peak energy around 10~MeV. Typically, $\tau_{\phi}=10^{12}$ s with $f_\phi\gtrsim10$ induces the neutrino flux that exceeds the observed values.

According to the observed neutrino flux, an upper limit on $\tau_\phi$ for certain $f_\phi$ can be obtained by using the binned statistical analysis with  the Poisson likelihood function~\cite{Ding:2018jdk, IceCube:2014rwe, Dev:2016uxj},
\begin{equation}
	L=\prod \limits_{i}\frac{e^{-n_i^{\rm th}}\left(n_i^{\rm th}\right)^{n_i^{\rm obs}}}{n_i^{\rm obs}!},
\end{equation}
where $n_i^{\rm th,obs}$ are the theory and observed values in the $i$-th bin, respectively.  A test statistic is then constructed as
\begin{equation}
	-2\Delta \ln L=-2\left(\ln L-\ln L_{\rm max}\right),
\end{equation}
where $\ln L_{\rm max}$ is the likelihood value with the observed neutrino flux by experiments.
The exclusion limit  at 90\% C.L. is then obtained with $-2\Delta \ln L=2.71$. The derived limit with $m_{\chi}=10$ GeV, $m_{\phi}=50$~GeV is the dashed black line in Fig. \ref{FIG:CSM}.
The limit from neutrino flux is weaker than those from CMB and BBN when $\tau_{\phi}\lesssim10^{15}$ s, otherwise it becomes the most stringent one at present. Notably the derived limit from neutrino flux also depends on masses of DM and dark scalar. So in the following numerical scan, we calculate the likelihood value for each sample to determine whether it is excluded. We find that under the constraints from $N_\text{eff}$, CMB and BBN, the neutrino flux can not exclude any samples individually. Therefore, no samples sprayed black are shown in Fig. \ref{FIG:CSM}.

\section{Collider Signature} \label{SEC:CSN} 

In this section, we consider the promising signatures of sterile neutrino and dark scalar at colliders. For sterile neutrinos in the mass range of 1 GeV$\lesssim m_N\lesssim m_W$, the most promising one is the displaced vertex signature. Via the Higgs portal coupling $\lambda_{3} \phi^2 (\Phi^\dag \Phi)$, the dark scalar could induce large Higgs invisible decay for certain $\lambda_3$.

\subsection{Displaced Vertex Signature}

\begin{figure}
		\begin{center}
			\includegraphics[width=0.45\linewidth]{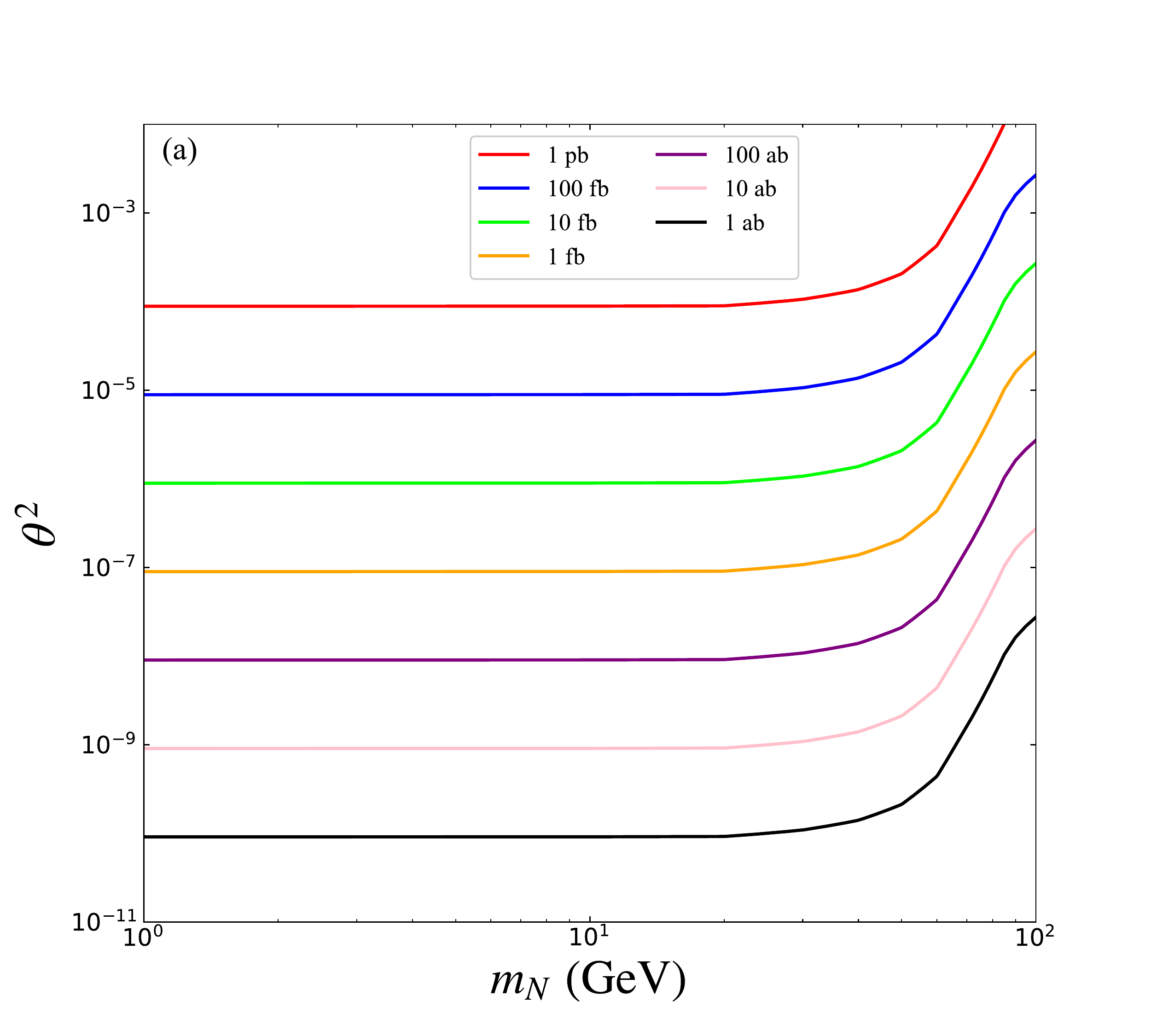}
			\includegraphics[width=0.45\linewidth]{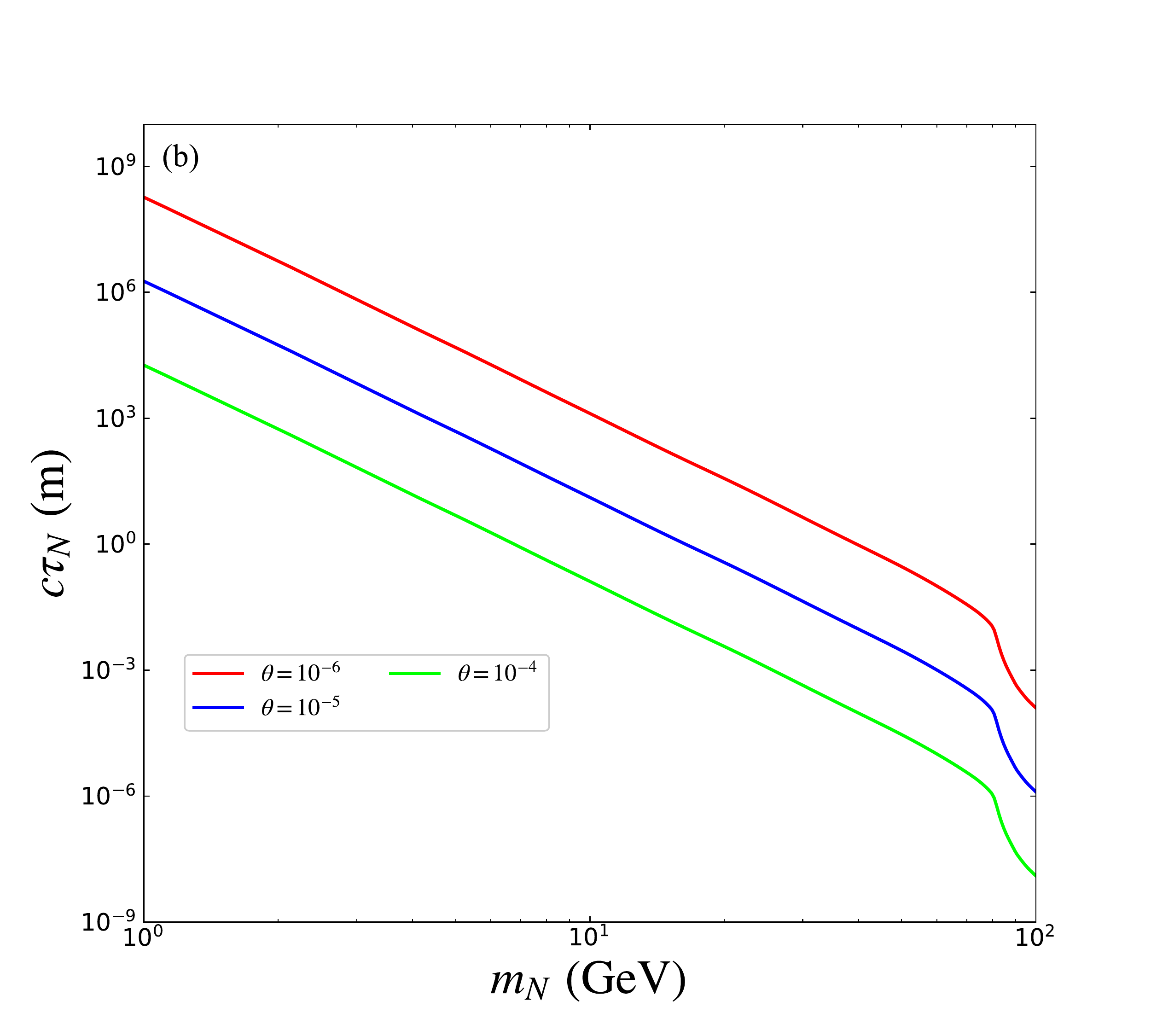}
		\end{center}
		\caption{Left: Cross section of $pp\to \ell^\pm N$  at the 14 TeV LHC. Right: Decay length of sterile neutrino. }
		\label{FIG:DisVert}
	\end{figure}

Within the framework of the type-I seesaw, the collider phenomenology of sterile neutrino is	determined by the mixing angle $\theta$ and mass $m_N$. The extensively studied production process is $pp\to W\to \ell^\pm N$ at LHC. In panel (a) of Fig. \ref{FIG:DisVert}, we show the production cross section $\sigma(pp\to \ell^\pm N)$ at the 14 TeV LHC. For sterile neutrino lighter than about 30 GeV, the production cross section is approximately a constant, which is proportional to $\theta^2$. For instance, $\theta^2\simeq10^{-5}(10^{-9})$ leads to the cross section at the order of 100 fb (10~ab). Therefore, when the mixing angle $\theta$ is too small, e.g. $\theta^2\lesssim10^{-10}$, the expected signal event will be less than ten even with an ultimate integrated luminosity of $3000~\text{fb}^{-1}$ and 100\% cut efficiency. For sterile neutrino heavier than 40 GeV, the phase space suppression by $m_N$ becomes obvious too.

The sterile neutrino decays into the SM quarks and leptons via the off-shell $W/Z$ boson. The explicit partial decay widths can be found in Ref. \cite{Jana:2018rdf}, and the total decay width can be  estimated as \cite{Drewes:2019fou}
\begin{equation}
	\Gamma_N\simeq 11.9\times \frac{G_F^2}{96\pi^3}\theta^2 m_N^5.
\end{equation}
In panel (b) of Fig. \ref{FIG:DisVert}, we show the decay length $c\tau_N$. For sterile neutrino lighter than 1 GeV, the decay length is usually too large to probe. Meanwhile, a sterile neutrino will decay promptly when it is heavier than $W$ or with a relatively large $\theta$. The promising region of displaced vertex signature is thus around 10 GeV scale with proper $\theta$ value.

Currently, searches for sterile neutrinos with displaced vertex signature have been performed at LHC \cite{ATLAS:2019kpx,CMS:2022fut,ATLAS:2022atq}. The exclusion limit could reach $m_N\lesssim15$ GeV and $\theta^2\gtrsim 3.6\times10^{-7}$ \cite{CMS:2022fut}. In the future, several experiments such as SHiP~\cite{SHiP:2018xqw, Gorbunov:2020rjx}, CEPC~\cite{CEPCStudyGroup:2018ghi}, LHC~\cite{Pascoli:2018heg, Izaguirre:2015pga} and FCC-hh~\cite{Antusch:2016ejd} could detect the parameter space with $m_N\lesssim m_W$ and $\theta^2\gtrsim10^{-11}$. These limits can be found in Fig. \ref{FIG:CLD}, however, none of them could reach the theoretical seesaw limit \cite{Esteban:2020cvm}.

\subsection{Higgs Invisible Decay}

As discussed in the previous section, the dark scalar $\phi$ is also long-lived. Anyway, the final states of the delayed decay $\phi\to \chi\nu$ are both invisible at colliders. Therefore, the process $h\to \phi\phi$ with $\phi\to \chi\nu$ contributes to the SM Higgs invisible decay. The corresponding decay width is
\begin{equation}
	\Gamma_{h\to\phi\phi}=\frac{\lambda_3^2v^2}{8\pi m_h}\sqrt{1-\frac{4m_\phi^2}{m_h^2}}.
\end{equation}

\begin{figure}
	\begin{center}
		\includegraphics[width=0.45\linewidth]{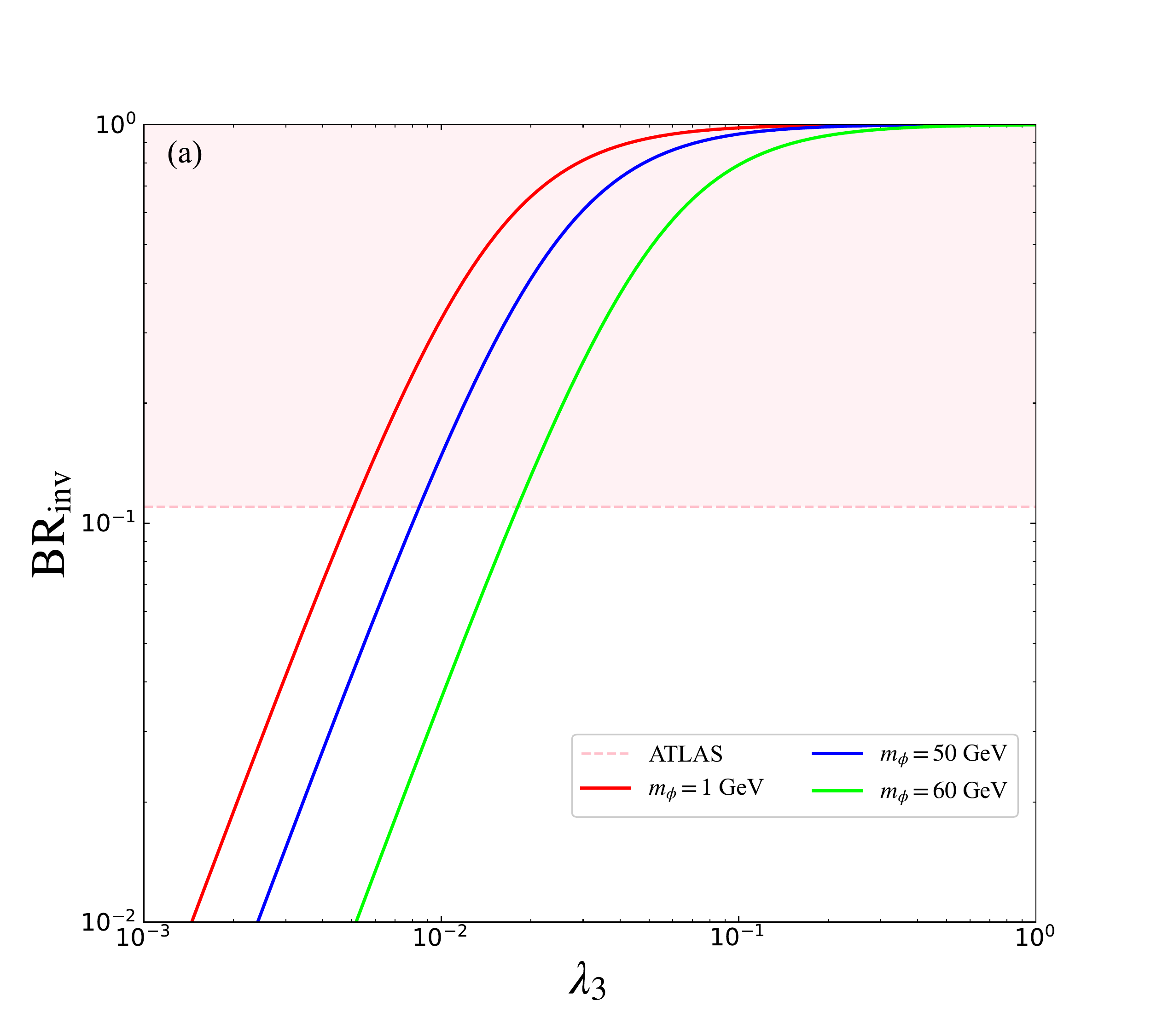}
		\includegraphics[width=0.45\linewidth]{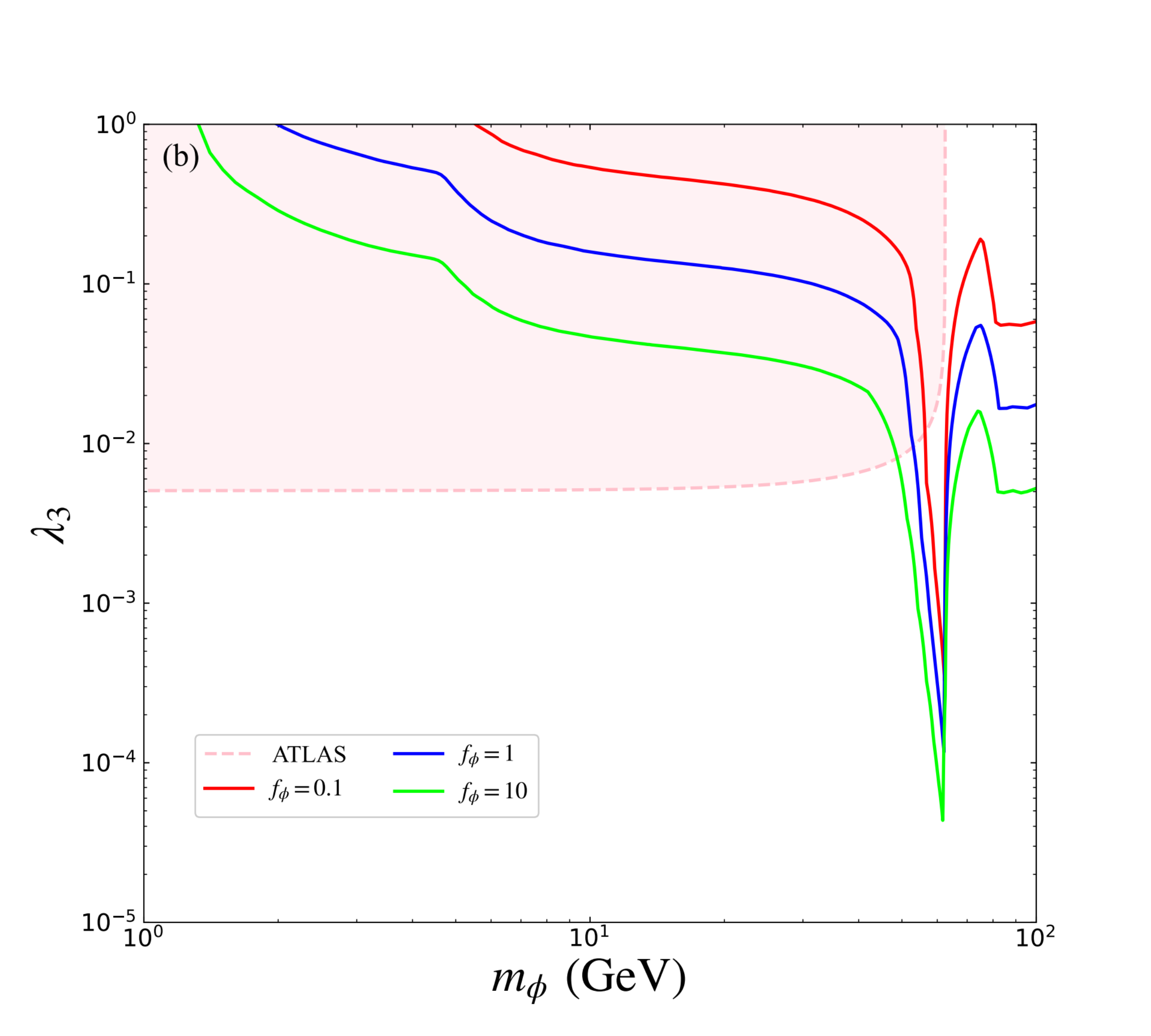}
	\end{center}
	\caption{Higgs invisible decay in the WIMP scalar scenario. The shaded region BR$_{\rm inv}>0.11$ is excluded by the ATLAS experiment \cite{ATLAS:2020kdi}. }
	\label{FIG:HigInv}
\end{figure}

The invisible branching ratio is calculated as BR$_{\rm inv}=\Gamma_{h\to\phi\phi}/(\Gamma_{h\to\phi\phi}+\Gamma_{\rm SM})$, with $\Gamma_{\rm SM}\approx4$ MeV. In the FIMP scalar scenario, $\lambda_{3}\lesssim10^{-11}$ leads to a neglectable contribution to Higgs invisible decay, so we only consider the WIMP scalar scenario. The predicted invisible branching ratio for certain $m_\phi$ is shown in panel (a) of Fig.~\ref{FIG:HigInv}. It is clear that the current measurement of invisible decay favors the region $\lambda_{3}\lesssim0.01$. However, according to the results in panel (b) of Fig.~\ref{FIG:DM}, such small $\lambda_{3}$ will lead the DM relic density over abundance. In panel (b) of Fig.~\ref{FIG:HigInv}, we have required that the correct relic density is obtained via the $\phi\to \chi\nu$ decay. It indicates that the mass region $m_\phi\lesssim50$ GeV is excluded.

\section{Combined Results}\label{SEC:CB}

\begin{figure}
	\begin{center}
		\includegraphics[width=0.45\linewidth]{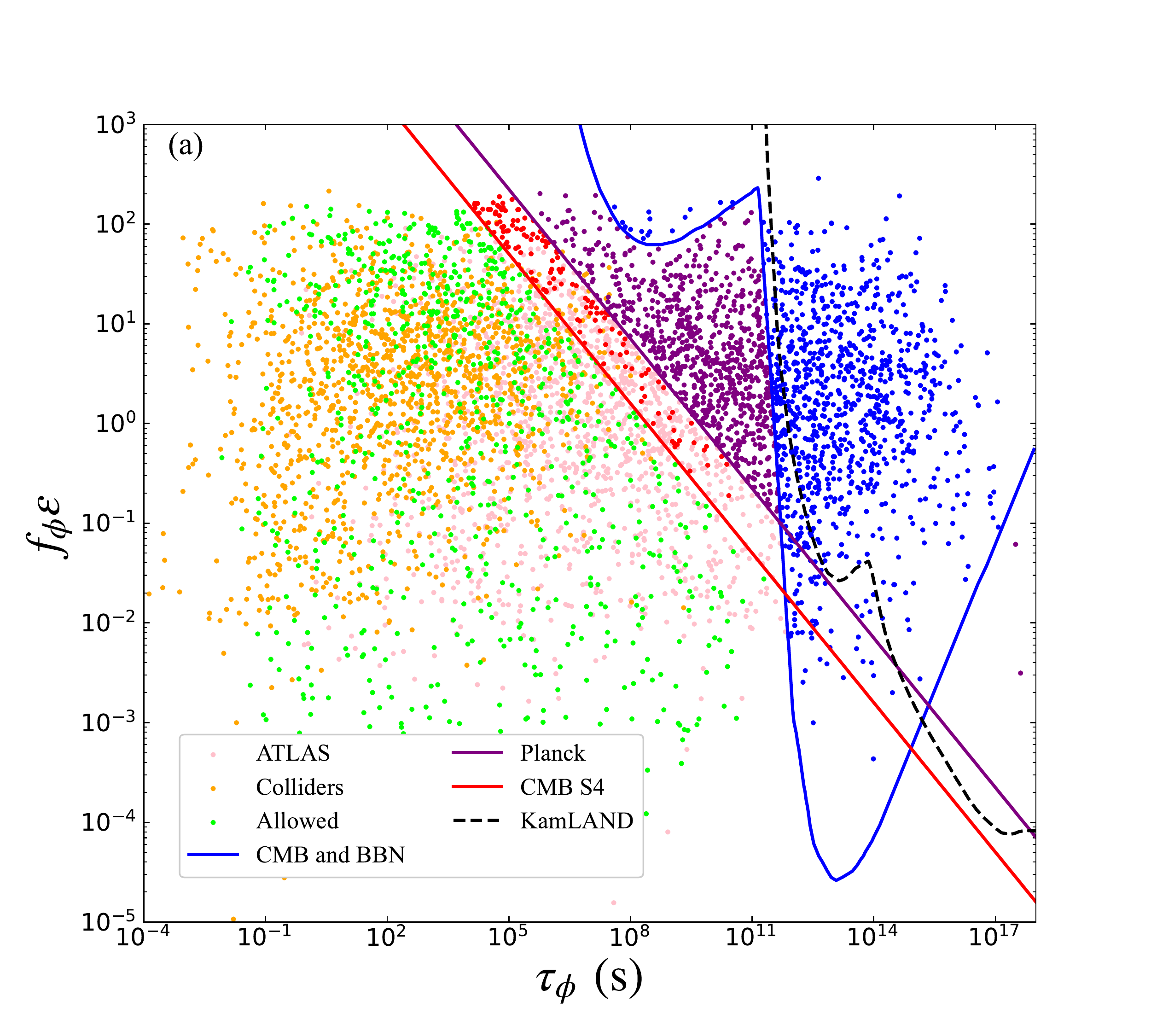}
		\includegraphics[width=0.45\linewidth]{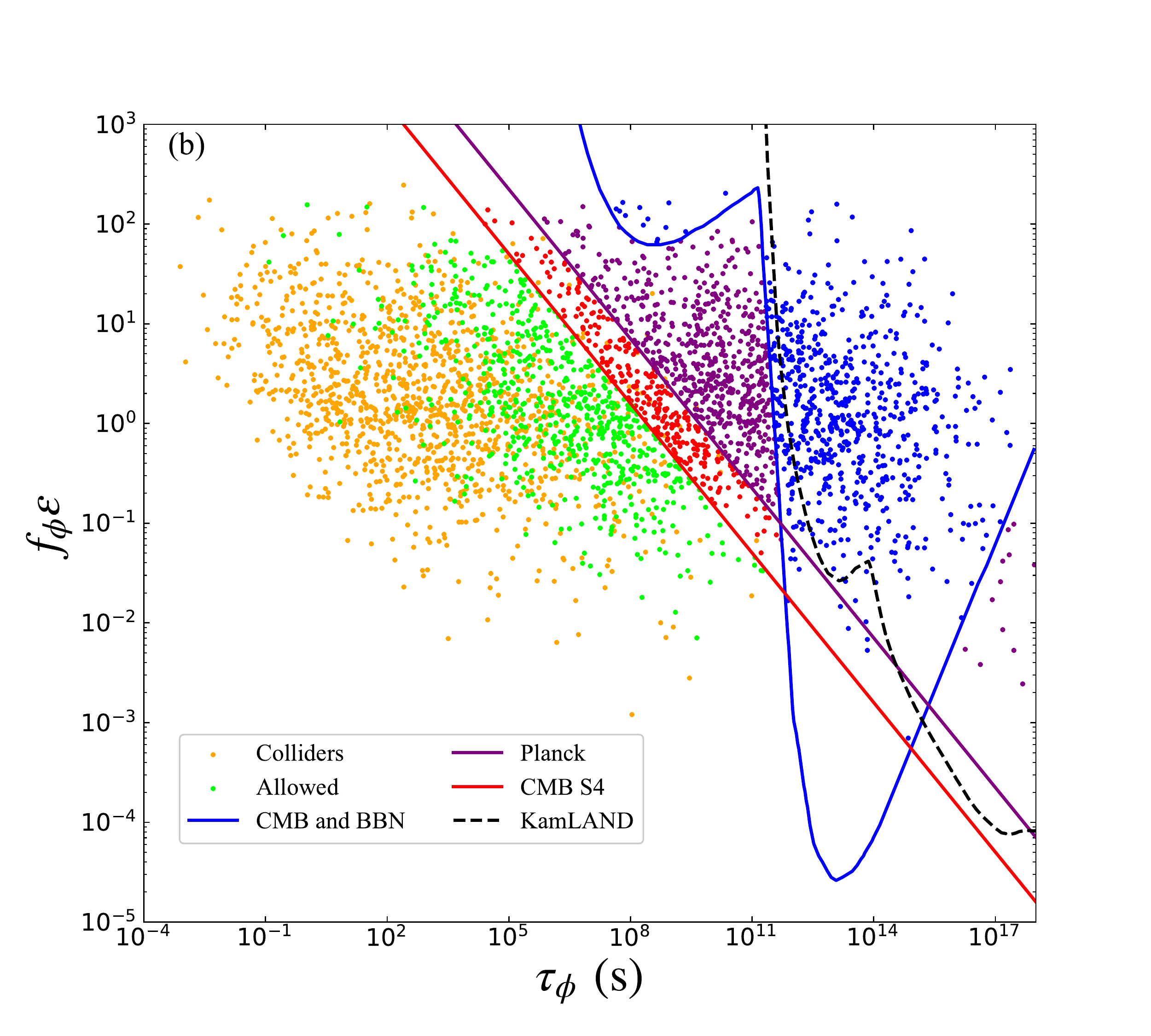}
	\end{center}
	\caption{Combined analysis for WIMP scalar scenario (left) and FIMP scalar scenario (right).  The orange points are excluded by the current direct search for sterile neutrino at colliders \cite{Abdullahi:2022jlv}. The pink samples (only in the left panel) are excluded by ATLAS measurement on Higgs invisible decay \cite{ATLAS:2020kdi}. The blue dots are excluded by CMB and BBN~\cite{Hambye:2021moy}. The purple samples are limited by current Planck $N_{\rm eff}$ observations~\cite{Planck:2018vyg}. The red points are within the reach of the future CMB S4~\cite{Abazajian:2019eic}. The black dotted line represents the upper limit obtained by  neutrino fluxes at KamLAND~\cite{KamLAND:2011bnd} and SK~\cite{Super-Kamiokande:2013ufi}. The green dots indicate the final allowed parameter space. }
	\label{FIG:CSM}
\end{figure}

As discussed in previous sections, the cosmological observables set constraints on the relatively small mixing angle, while the collider signatures favor a large mixing angle. In this section, we combine the individual constraint to obtain the viable parameter space. The free parameter set is \{$m_N, m_\phi, m_\chi, \lambda_3, \lambda_{ds}, \theta$\} in this study. We perform a random scan in the following range
\begin{eqnarray}
	\begin{aligned}
	 m_N\in[1,100]~\GeV, m_{\phi,\chi}\in[0.1,100]~\GeV,  \lambda_{ds}\in[10^{-13},10^{-10}], \theta\in[10^{-7},1].
	\end{aligned}
\end{eqnarray}
As for the Higgs portal coupling, we consider 
$\lambda_{3}\in[0.001,1]$ for the WIMP scalar scenario and $\lambda_{3}\in[10^{-13},10^{-10}]$ for the FIMP scalar scenario. During the scan, we require the relic density satisfies the Planck observed result in the $3\sigma$ range \cite{Planck:2018vyg}, i.e., $\Omega_\chi h^2\in[0.117,0.123]$. To obtain the specific scenario $m_N>m_\phi+m_\chi$ and $m_\phi>m_\chi$, these conditions are also imposed manually.

\begin{figure}
	\begin{center}
		\includegraphics[width=0.45\linewidth]{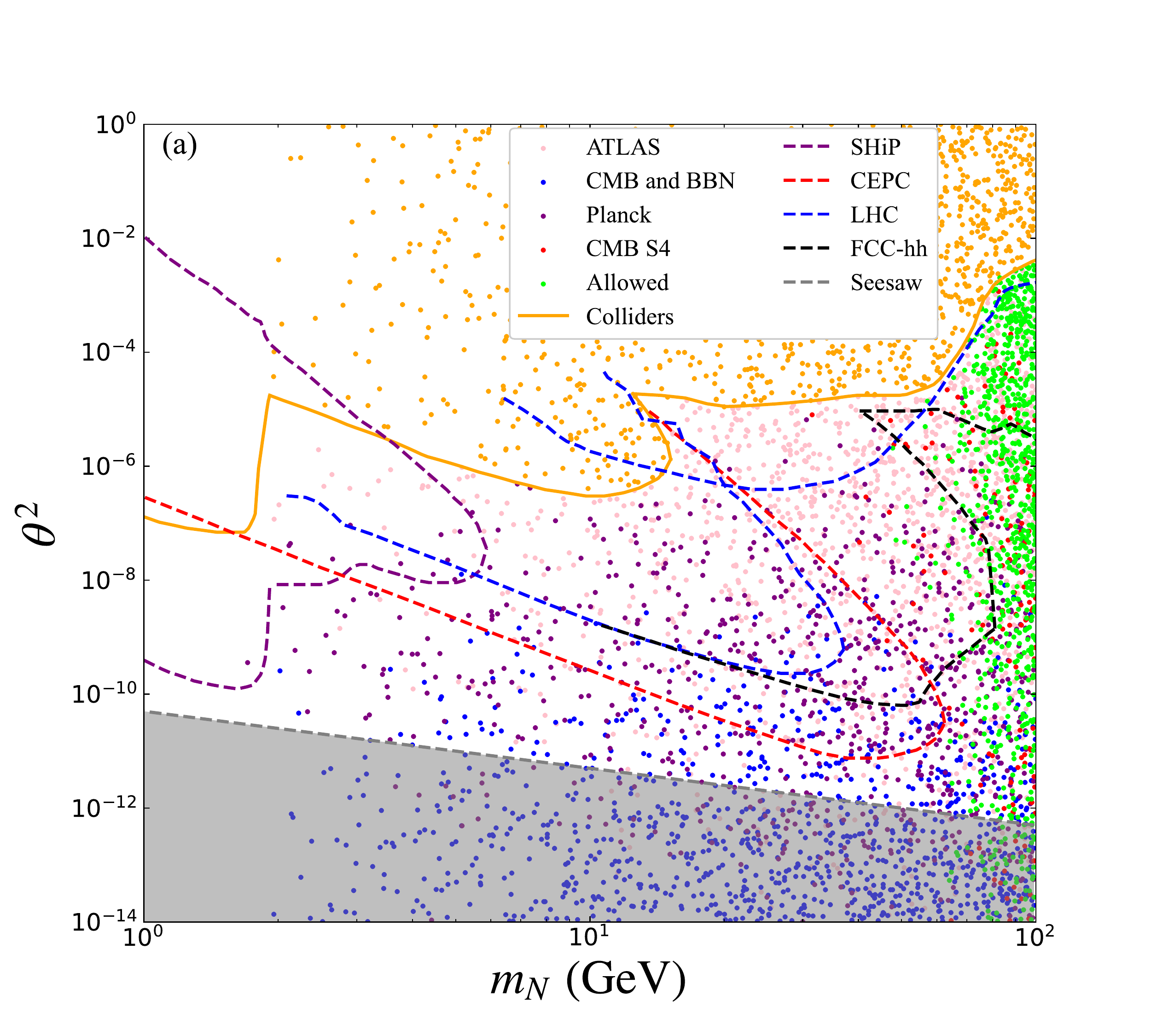}
		\includegraphics[width=0.45\linewidth]{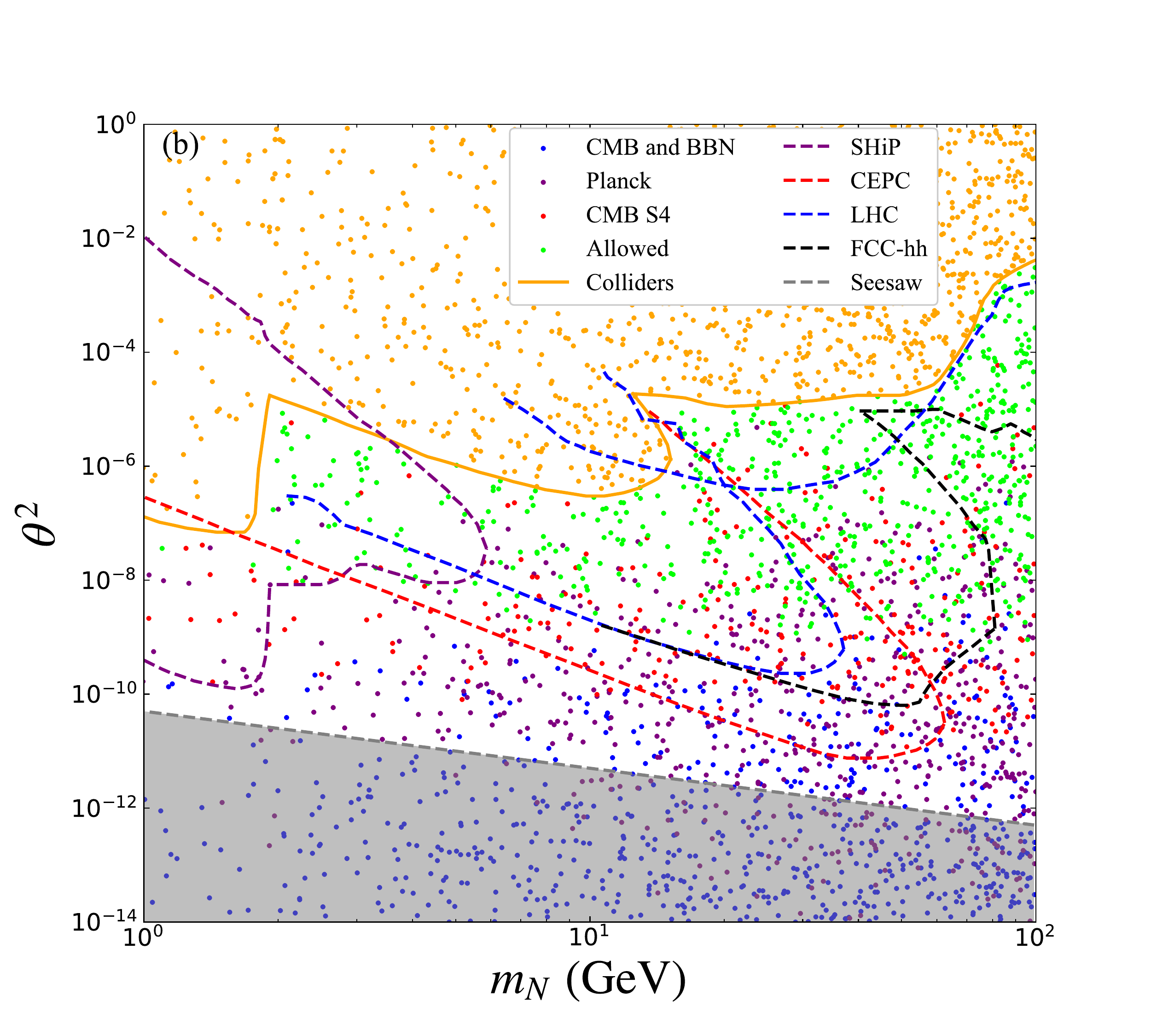}
	\end{center}
	\caption{Constraints on the heavy sterile neutrino for the WIMP scalar scenario (left) and the FIMP scalar scenario (right). The orange line is the current exclusion limit for sterile neutrino at colliders \cite{Abdullahi:2022jlv}. The purple, red, blue and black dashed lines are the future limits from SHiP~\cite{SHiP:2018xqw, Gorbunov:2020rjx}, CEPC~\cite{CEPCStudyGroup:2018ghi}, LHC~\cite{Pascoli:2018heg, Izaguirre:2015pga} and FCC-hh~\cite{Antusch:2016ejd}, respectively. The gray solid line indicates the seesaw predicted mixing angle with Eq.~\eqref{Eqn:SS}. Other labels are the same as Fig.~\ref{FIG:CSM}. }
	\label{FIG:CLD}
\end{figure}

The combined results in the $f_\phi\varepsilon-\tau_\phi$ plane are shown in Fig.~\ref{FIG:CSM}. Due to the massive DM in the decay of $\phi$, the fractional abundance $f_\phi$ in this analysis needs to be multiplied by $\varepsilon=(m^2_{\phi}-m^2_{\chi})/2m^2_{\phi}$ to keep consistent with the constraints in Ref. \cite{Hambye:2021moy}. The cosmological constraints from CMB and BBN exclude the parameter space with $\tau_{\phi}\gtrsim10^{12}$ s. Meanwhile, the Planck $N_\text{eff}$ result can exclude $\tau_{\phi}\gtrsim10^5$ s with $f_\phi\varepsilon\lesssim10^2$. Recall that $\lambda_{3}\sim\mathcal{O}(1)$ could lead to $Y_\phi$ much smaller than $Y_\chi$, so the distribution of $f_\phi\varepsilon$ for the WIMP scalar scenario is wider than the FIMP scalar scenario. It is obvious that the limit from colliders tends to exclude dark scalar with short lifetime for the FIMP scalar scenario, and the allowed parameter space tends to the area with larger $\tau_\phi$ when $f_\phi\varepsilon$ decreases.

To figure out the promising region at colliders, we then show the scanned samples in the $\theta^2-m_N$ plane in Fig.~\ref{FIG:CLD}. Panel (a) is the result for the WIMP scalar scenario. We can see that $m_N\lesssim2$ GeV is prohibited, because a too light $m_\phi$ will cause the over production of DM $\chi$. Although future colliders can cover most parameter space in the $\theta^2-m_N$ plane, we find that the constraint from Higgs invisible decay has already excluded the displaced vertex signature promising region at SHiP, CEPC and LHC. There are a few samples around the range of 50 GeV to 80 GeV with $\theta^2\in[10^{-9},10^{-6}]$ that can still be detected by future FCC-hh with the displaced signal. Such region may be double checked by the CMB S4 experiment via the measurement of $N_\text{eff}$.

As shown in panel (b) of Fig.~\ref{FIG:CLD}, the situation for the FIMP scalar scenario is quite different from the WIMP one. Without the constraint from Higgs invisible decay, $m_N$ can down to the scanned lower limit of 1 GeV. So once a sterile neutrino with a mass less than 50 GeV is found, only the FIMP scalar scenario is favored. Meanwhile, cosmological constraints from CMB, BBN and $N_{\rm eff}$ have more clear lower limits on allowed samples ($\theta^2\gtrsim10^{-10}$), which implies that these lower limits on $\theta^2$ are already about one order of magnitude higher than the minimal seesaw prediction. In this way, we definitely expect observable displaced vertex signature at future colliders. There is a small mass gap between 2 GeV and 10 GeV that is hard to detect at colliders, but is within the reach of CMB S4 measurement on $N_\text{eff}$. Otherwise, if no clear $\Delta N_\text{eff}$ is observed at CMB S4, we should have $\theta^2\gtrsim10^{-9}$.

\section{Conclusion} \label{SEC:CL} 
The sterile neutrino portal dark matter model is an appealing pathway to explain the origin of tiny neutrino mass and dark matter simultaneously. Besides the sterile neutrino $N$, a dark sector with one fermion singlet $\chi$ and one scalar singlet $\phi$ is also introduced. In this paper, we consider the fermion singlet $\chi$ as FIMP dark matter. Although the DM itself is hard to detect, the dark scalar $\phi$ and sterile neutrino $N$ could lead to observable signatures at cosmology and colliders. 

We focus on the specific scenario $m_N>m_\phi+m_\chi$, where the decay of dark scalar $\phi\to\chi\nu$ is heavily suppressed. Such delayed decay will have a great impact on cosmological observables such as the Big Bang Nucleosynthesis, the Cosmic Microwave Background anisotropy power spectra, the effective number of relativistic neutrino species $N_{\rm eff}$ and the energetic neutrino flux. The cosmological constraints from CMB and BBN exclude the parameter space with $\tau_{\phi}\gtrsim10^{12}$ s. Meanwhile, the Planck measurement of $N_\text{eff}$ sets the most stringent constraint for $\tau_{\phi}\lesssim10^{12}$ s. We also find that restriction from neutrino flux is weaker than those from CMB and BBN.

The sterile neutrino $N$ can be also long-lived when lighter than $W$ boson, which then generates a displaced vertex signature at colliders. Currently, the mixing angle with $\theta^2\gtrsim10^{-6}$ is excluded. In the future, parameter space with $\theta^2\gtrsim10^{-10}$ could be detected. But this limit is still higher than the minimal seesaw prediction. In the WIMP scalar scenario, the dark scalar induces Higgs invisible decay via the quartic coupling $\lambda_3 \phi^2(\Phi^\dag \Phi)$. By requiring the correct relic density of DM, we find that $m_\phi\lesssim50$ GeV is already excluded by Higgs invisible decay.

Since the constraints from cosmology and colliders are complementary to each other, a random scan over parameter space is also performed. For the WIMP scalar scenario, we find that under tight constraint from Higgs invisible decay, only the FCC-hh has the potential to probe the sterile neutrino around 50 GeV to 80 GeV. For the FIMP scalar scenario, the sterile neutrino can be as light as 1 GeV, thus displaced vertex signatures are expected at all colliders. We also find that current cosmological constraints on $\theta^2$ are already higher than the minimal seesaw limit. So the future CMB S4 measurement of $N_\text{eff}$ together with displaced vertex searches at colliders are able to cover almost the whole parameter space with $m_N\lesssim m_W$.

\section*{Acknowledgments}
This work is supported by the National Natural Science Foundation of China under Grant No. 11975011, 11805081 and  11635009, Natural Science Foundation of Shandong Province under Grant No. ZR2019QA021 and ZR2018MA047.



\begin{thebibliography}{000} 
\bibitem{Krauss:2002px}
L.~M.~Krauss, S.~Nasri and M.~Trodden,
Phys. Rev. D \textbf{67}, 085002 (2003)
[arXiv:hep-ph/0210389 [hep-ph]].

\bibitem{Asaka:2005an}
T.~Asaka, S.~Blanche and M.~Shaposhnikov,
Phys. Lett. B \textbf{631}, 151-156 (2005)
[arXiv:hep-ph/0503065 [hep-ph]].

\bibitem{Ma:2006km}
E.~Ma,
Phys. Rev. D \textbf{73}, 077301 (2006)
[arXiv:hep-ph/0601225 [hep-ph]].

\bibitem{Aoki:2008av}
M.~Aoki, S.~Kanemura and O.~Seto,
Phys. Rev. Lett. \textbf{102}, 051805 (2009)
[arXiv:0807.0361 [hep-ph]].


\bibitem{Cai:2017jrq}
Y.~Cai, J.~Herrero-Garc\'\i{}a, M.~A.~Schmidt, A.~Vicente and R.~R.~Volkas,
Front. in Phys. \textbf{5}, 63 (2017)
[arXiv:1706.08524 [hep-ph]].


\bibitem{Minkowski:1977sc}
P.~Minkowski,
Phys.\ Lett.\ B {\bf 67}, 421 (1977).

\bibitem{Mohapatra:1979ia}
R.~N.~Mohapatra and G.~Senjanovic,
Phys.\ Rev.\ Lett.\  {\bf 44}, 912 (1980).

\bibitem{Fukugita:1986hr}
M.~Fukugita and T.~Yanagida,
Phys. Lett. B \textbf{174}, 45-47 (1986)

\bibitem{Dasgupta:2021ies}
B.~Dasgupta and J.~Kopp,
Phys. Rept. \textbf{928}, 1-63 (2021)
[arXiv:2106.05913 [hep-ph]].

\bibitem{Abdullahi:2022jlv}
A.~M.~Abdullahi, P.~B.~Alzas, B.~Batell, A.~Boyarsky, S.~Carbajal, A.~Chatterjee, J.~I.~Crespo-Anadon, F.~F.~Deppisch, A.~De Roeck and M.~Drewes, \textit{et al.}
[arXiv:2203.08039 [hep-ph]].

\bibitem{Gorbunov:2007ak}
D.~Gorbunov and M.~Shaposhnikov,
JHEP \textbf{10}, 015 (2007)
[erratum: JHEP \textbf{11}, 101 (2013)]
[arXiv:0705.1729 [hep-ph]].

\bibitem{Atre:2009rg}
A.~Atre, T.~Han, S.~Pascoli and B.~Zhang,
JHEP \textbf{05}, 030 (2009)
[arXiv:0901.3589 [hep-ph]].

\bibitem{Deppisch:2015qwa}
F.~F.~Deppisch, P.~S.~Bhupal Dev and A.~Pilaftsis,
New J. Phys. \textbf{17}, no.7, 075019 (2015)
[arXiv:1502.06541 [hep-ph]].

\bibitem{Cai:2017mow}
Y.~Cai, T.~Han, T.~Li and R.~Ruiz,
Front. in Phys. \textbf{6}, 40 (2018)
[arXiv:1711.02180 [hep-ph]].

\bibitem{Helo:2013esa}
J.~C.~Helo, M.~Hirsch and S.~Kovalenko,
Phys. Rev. D \textbf{89}, 073005 (2014)
[erratum: Phys. Rev. D \textbf{93}, no.9, 099902 (2016)]
[arXiv:1312.2900 [hep-ph]].

\bibitem{Alimena:2019zri}
J.~Alimena, J.~Beacham, M.~Borsato, Y.~Cheng, X.~Cid Vidal, G.~Cottin, A.~De Roeck, N.~Desai, D.~Curtin and J.~A.~Evans, \textit{et al.}
J. Phys. G \textbf{47}, no.9, 090501 (2020)
[arXiv:1903.04497 [hep-ex]].

\bibitem{Escudero:2016tzx}
M.~Escudero, N.~Rius and V.~Sanz,
JHEP \textbf{02}, 045 (2017)
[arXiv:1606.01258 [hep-ph]].

\bibitem{Escudero:2016ksa}
M.~Escudero, N.~Rius and V.~Sanz,
Eur. Phys. J. C \textbf{77}, no.6, 397 (2017)
[arXiv:1607.02373 [hep-ph]].

\bibitem{Coito:2022kif}
L.~Coito, C.~Faubel, J.~Herrero-Garc\'\i{}a, A.~Santamaria and A.~Titov,
JHEP \textbf{08}, 085 (2022)
[arXiv:2203.01946 [hep-ph]].

\bibitem{Coy:2022xfj}
R.~Coy and A.~Gupta,
[arXiv:2211.05091 [hep-ph]].

\bibitem{Li:2022xjx}
S.~P.~Li and X.~J.~Xu,
[arXiv:2212.09109 [hep-ph]].

\bibitem{Campos:2017odj}
M.~D.~Campos, F.~S.~Queiroz, C.~E.~Yaguna and C.~Weniger,
JCAP \textbf{07}, 016 (2017)
[arXiv:1702.06145 [hep-ph]].

\bibitem{Batell:2017rol}
B.~Batell, T.~Han and B.~Shams Es Haghi,
Phys. Rev. D \textbf{97}, no.9, 095020 (2018)
[arXiv:1704.08708 [hep-ph]].


\bibitem{Folgado:2018qlv}
M.~G.~Folgado, G.~A.~G\'omez-Vargas, N.~Rius and R.~Ruiz De Austri,
JCAP \textbf{08}, 002 (2018)
[arXiv:1803.08934 [hep-ph]].

\bibitem{Bandyopadhyay:2020qpn}
P.~Bandyopadhyay, E.~J.~Chun and R.~Mandal,
JCAP \textbf{08}, 019 (2020)
[arXiv:2005.13933 [hep-ph]].

\bibitem{Cheng:2020gut}
Y.~Cheng and W.~Liao,
Phys. Lett. B \textbf{815}, 136118 (2021)
[arXiv:2012.01875 [hep-ph]].

\bibitem{Falkowski:2017uya}
A.~Falkowski, E.~Kuflik, N.~Levi and T.~Volansky,
Phys. Rev. D \textbf{99}, no.1, 015022 (2019)
[arXiv:1712.07652 [hep-ph]].

\bibitem{Liu:2020mxj}
A.~Liu, Z.~L.~Han, Y.~Jin and F.~X.~Yang,
Phys. Rev. D \textbf{101}, no.9, 095005 (2020)
[arXiv:2001.04085 [hep-ph]].

\bibitem{Chang:2021ose}
Z.~F.~Chang, Z.~X.~Chen, J.~S.~Xu and Z.~L.~Han,
JCAP \textbf{06}, 006 (2021)
[arXiv:2104.02364 [hep-ph]].

\bibitem{Adams:1998nr}
J.~A.~Adams, S.~Sarkar and D.~W.~Sciama,
Mon. Not. Roy. Astron. Soc. \textbf{301}, 210-214 (1998)
[arXiv:astro-ph/9805108 [astro-ph]].

\bibitem{Chen:2003gz}
X.~L.~Chen and M.~Kamionkowski,
Phys. Rev. D \textbf{70}, 043502 (2004)
[arXiv:astro-ph/0310473 [astro-ph]].

\bibitem{Boyarsky:2021yoh}
A.~Boyarsky, M.~Ovchynnikov, N.~Sabti and V.~Syvolap,
Phys. Rev. D \textbf{104}, no.3, 035006 (2021)
[arXiv:2103.09831 [hep-ph]].

\bibitem{Liu:2022rst}
A.~Liu, F.~L.~Shao, Z.~L.~Han, Y.~Jin and H.~Li,
[arXiv:2205.11846 [hep-ph]].

\bibitem{Barman:2022scg}
B.~Barman, P.~S.~B.~Dev and A.~Ghoshal,
[arXiv:2210.07739 [hep-ph]].

\bibitem{Hambye:2021moy}
T.~Hambye, M.~Hufnagel and M.~Lucca,
JCAP \textbf{05}, no.05, 033 (2022)
[arXiv:2112.09137 [hep-ph]].

\bibitem{Coy:2021sse}
R.~Coy, A.~Gupta and T.~Hambye,
Phys. Rev. D \textbf{104}, no.8, 083024 (2021)
[arXiv:2104.00042 [hep-ph]].

\bibitem{Mohapatra:1986bd}
R.~N.~Mohapatra and J.~W.~F.~Valle,
Phys. Rev. D \textbf{34}, 1642 (1986)

\bibitem{Mohapatra:1986aw}
R.~N.~Mohapatra,
Phys. Rev. Lett. \textbf{56}, 561-563 (1986)

\bibitem{Wyler:1982dd}
D.~Wyler and L.~Wolfenstein,
Nucl. Phys. B \textbf{218}, 205-214 (1983)

\bibitem{Akhmedov:1995ip}
E.~K.~Akhmedov, M.~Lindner, E.~Schnapka and J.~W.~F.~Valle,
Phys. Lett. B \textbf{368}, 270-280 (1996)
[arXiv:hep-ph/9507275 [hep-ph]].

\bibitem{Akhmedov:1995vm}
E.~K.~Akhmedov, M.~Lindner, E.~Schnapka and J.~W.~F.~Valle,
Phys. Rev. D \textbf{53}, 2752-2780 (1996)
[arXiv:hep-ph/9509255 [hep-ph]].

\bibitem{Belanger:2013oya}
G.~Belanger, F.~Boudjema, A.~Pukhov and A.~Semenov,
Comput. Phys. Commun. \textbf{185}, 960-985 (2014)
[arXiv:1305.0237 [hep-ph]].


\bibitem{Planck:2018vyg}
N.~Aghanim \textit{et al.} [Planck],
Astron. Astrophys. \textbf{641}, A6 (2020)
[erratum: Astron. Astrophys. \textbf{652}, C4 (2021)]
[arXiv:1807.06209 [astro-ph.CO]].

\bibitem{EscuderoAbenza:2020cmq}
M.~Escudero Abenza,
JCAP \textbf{05}, 048 (2020)
[arXiv:2001.04466 [hep-ph]].

\bibitem{Escudero:2018mvt}
M.~Escudero,
JCAP \textbf{02}, 007 (2019)
[arXiv:1812.05605 [hep-ph]].

\bibitem{Abazajian:2019eic}
K.~Abazajian, G.~Addison, P.~Adshead, Z.~Ahmed, S.~W.~Allen, D.~Alonso, M.~Alvarez, A.~Anderson, K.~S.~Arnold and C.~Baccigalupi, \textit{et al.}
[arXiv:1907.04473 [astro-ph.IM]].

\bibitem{Blackadder:2014wpa}
G.~Blackadder and S.~M.~Koushiappas,
Phys. Rev. D \textbf{90}, no.10, 103527 (2014)
[arXiv:1410.0683 [astro-ph.CO]].

\bibitem{Mangano:2005cc}
G.~Mangano, G.~Miele, S.~Pastor, T.~Pinto, O.~Pisanti and P.~D.~Serpico,
Nucl. Phys. B \textbf{729}, 221-234 (2005)
[arXiv:hep-ph/0506164 [hep-ph]].

\bibitem{Grohs:2015tfy}
E.~Grohs, G.~M.~Fuller, C.~T.~Kishimoto, M.~W.~Paris and A.~Vlasenko,
Phys. Rev. D \textbf{93}, no.8, 083522 (2016)
[arXiv:1512.02205 [astro-ph.CO]].

\bibitem{deSalas:2016ztq}
P.~F.~de Salas and S.~Pastor,
JCAP \textbf{07}, 051 (2016)
[arXiv:1606.06986 [hep-ph]].

\bibitem{Ding:2018jdk}
R.~Ding, Z.~L.~Han, L.~Huang and Y.~Liao,
Chin. Phys. C \textbf{42}, no.10, 103101 (2018)
[arXiv:1802.05248 [hep-ph]].

\bibitem{IceCube:2014rwe}
M.~G.~Aartsen \textit{et al.} [IceCube],
Phys. Rev. D \textbf{91}, no.2, 022001 (2015)
[arXiv:1410.1749 [astro-ph.HE]].

\bibitem{Dev:2016uxj}
P.~S.~B.~Dev, D.~K.~Ghosh and W.~Rodejohann,
Phys. Lett. B \textbf{762}, 116-123 (2016)
[arXiv:1605.09743 [hep-ph]].

\bibitem{Vitagliano:2019yzm}
E.~Vitagliano, I.~Tamborra and G.~Raffelt,
Rev. Mod. Phys. \textbf{92}, 45006 (2020)
[arXiv:1910.11878 [astro-ph.HE]].

\bibitem{KamLAND:2011bnd}
A.~Gando \textit{et al.} [KamLAND],
Astrophys. J. \textbf{745}, 193 (2012)
[arXiv:1105.3516 [astro-ph.HE]].

\bibitem{Super-Kamiokande:2013ufi}
H.~Zhang \textit{et al.} [Super-Kamiokande],
Astropart. Phys. \textbf{60}, 41-46 (2015)
[arXiv:1311.3738 [hep-ex]].

\bibitem{Super-Kamiokande:2015qek}
E.~Richard \textit{et al.} [Super-Kamiokande],
Phys. Rev. D \textbf{94}, no.5, 052001 (2016)
[arXiv:1510.08127 [hep-ex]].

\bibitem{Jana:2018rdf}
S.~Jana, N.~Okada and D.~Raut,
Phys. Rev. D \textbf{98}, no.3, 035023 (2018)
[arXiv:1804.06828 [hep-ph]].

\bibitem{Drewes:2019fou}
M.~Drewes and J.~Hajer,
JHEP \textbf{02}, 070 (2020)
[arXiv:1903.06100 [hep-ph]].

\bibitem{ATLAS:2019kpx}
G.~Aad \textit{et al.} [ATLAS],
JHEP \textbf{10}, 265 (2019)
[arXiv:1905.09787 [hep-ex]].

\bibitem{CMS:2022fut}
A.~Tumasyan \textit{et al.} [CMS],
JHEP \textbf{07}, 081 (2022)
[arXiv:2201.05578 [hep-ex]].

\bibitem{ATLAS:2022atq}
[ATLAS],
[arXiv:2204.11988 [hep-ex]].

\bibitem{SHiP:2018xqw}
C.~Ahdida \textit{et al.} [SHiP],
JHEP \textbf{04}, 077 (2019)
[arXiv:1811.00930 [hep-ph]].

\bibitem{Gorbunov:2020rjx}
D.~Gorbunov, I.~Krasnov, Y.~Kudenko and S.~Suvorov,
Phys. Lett. B \textbf{810}, 135817 (2020)
[arXiv:2004.07974 [hep-ph]].

\bibitem{CEPCStudyGroup:2018ghi}
J.~B.~Guimar\~aes da Costa \textit{et al.} [CEPC Study Group],
[arXiv:1811.10545 [hep-ex]].

\bibitem{Pascoli:2018heg}
S.~Pascoli, R.~Ruiz and C.~Weiland,
JHEP \textbf{06}, 049 (2019)
[arXiv:1812.08750 [hep-ph]].

\bibitem{Izaguirre:2015pga}
E.~Izaguirre and B.~Shuve,
Phys. Rev. D \textbf{91}, no.9, 093010 (2015)
[arXiv:1504.02470 [hep-ph]].

\bibitem{Antusch:2016ejd}
S.~Antusch, E.~Cazzato and O.~Fischer,
Int. J. Mod. Phys. A \textbf{32}, no.14, 1750078 (2017)
[arXiv:1612.02728 [hep-ph]].

\bibitem{Esteban:2020cvm}
I.~Esteban, M.~C.~Gonzalez-Garcia, M.~Maltoni, T.~Schwetz and A.~Zhou,
JHEP \textbf{09}, 178 (2020)
[arXiv:2007.14792 [hep-ph]].

\bibitem{ATLAS:2020kdi}
[ATLAS],
ATLAS-CONF-2020-052.

\end{thebibliography}
\end{document}